\documentclass[preprint]{elsarticle}
\usepackage{amsmath}
\usepackage{braket}
\newcommand*{\Scale}[2][4]{\scalebox{#1}{$#2$}}

\makeatletter
\renewcommand{\paragraph}{%
	\@startsection{paragraph}{4}%
	{\z@}{1.5ex \@plus 1ex \@minus 0.5ex}{-1em}%
	{\normalfont\normalsize\bfseries}%
}
\makeatother

\usepackage{tabularx}
\newcolumntype{Y}{>{\centering\arraybackslash}X}

\usepackage{amssymb}

\usepackage{algorithm}
\usepackage[noend]{algpseudocode}


\makeatletter
\renewcommand{\fnum@figure}{Fig. \thefigure}
\makeatother

\newtheorem{defn}{Definition}
\newtheorem{prop}{Proposition}
\newenvironment{proof}{\noindent \emph{Proof.}}{\qed}

\journal{Annals of Physics}
\begin{document}
\begin{frontmatter}
	\title{Quantum Algorithm for Testing Graph Completeness}
	
	\author[inst1]{Sara Giordano\corref{cor1}%
	}
	\ead{sgiordan@ucm.es}
	
	\author[inst1,inst2]{Miguel A. Martin-Delgado}
	
	\affiliation[inst1]{organization={Departamento de F\'{\i}sica Te\'orica, Universidad Complutense},
		city={Madrid},
		postcode={28040},
		country={Spain}}
	\affiliation[inst2]{organization={CCS-Center for Computational Simulation, Campus de Montegancedo Universidad Politecnica de Madrid (UPM)},
	city={Boadilla del Monte},
	postcode={28660, Madrid},
	country={Spain}}
	\cortext[cor1]{Corresponding author}
	
	\begin{abstract}
	Testing graph completeness is a critical problem in computer science and network theory. Leveraging quantum computation, we present an efficient algorithm using the Szegedy quantum walk and quantum phase estimation (QPE). Our algorithm, which takes the number of nodes and the adjacency matrix as input, constructs a quantum walk operator and applies QPE to estimate its eigenvalues. These eigenvalues reveal the graph's structural properties, enabling us to determine its completeness. We establish a relationship between the number of nodes in a complete graph and the number of marked nodes, optimizing the success probability and running time. The time complexity of our algorithm is $\mathcal{O}(\log^2n)$, where $n$ is the number of nodes of the graph. offering a clear quantum advantage over classical methods. This approach is useful in network structure analysis, evaluating classical routing algorithms, and assessing systems based on pairwise comparisons.
	\end{abstract}
	
	
	
%
%
%
	
\end{frontmatter}

	\section{Introduction and Main Results}\label{sec_I}

In recent decades, quantum computing has developed algorithms that provide computational advantages over classical methods \cite{Preskill_2012, NielsenAndChuang, Feynman_1982}, known as the theoretical quantum advantage, as opposed to the long-sought practical quantum advantage, currently in full pursuit of potential technological and commercial applications. These advancements span various fields, from solving complex mathematical problems like factoring large numbers \cite{Shor} and searching unsorted databases \cite{Grover, Grover_2, Ambainis_2014}, to simulating physical systems \cite{Seth_Lloyd_1996, NielsenAndChuang, RMP_Galindo_MA} and applications in computational biology \cite{Casares_2022, QQ_metropolis}.

\begin{figure}[t]
	\centering
	\includegraphics[width=0.5\columnwidth]{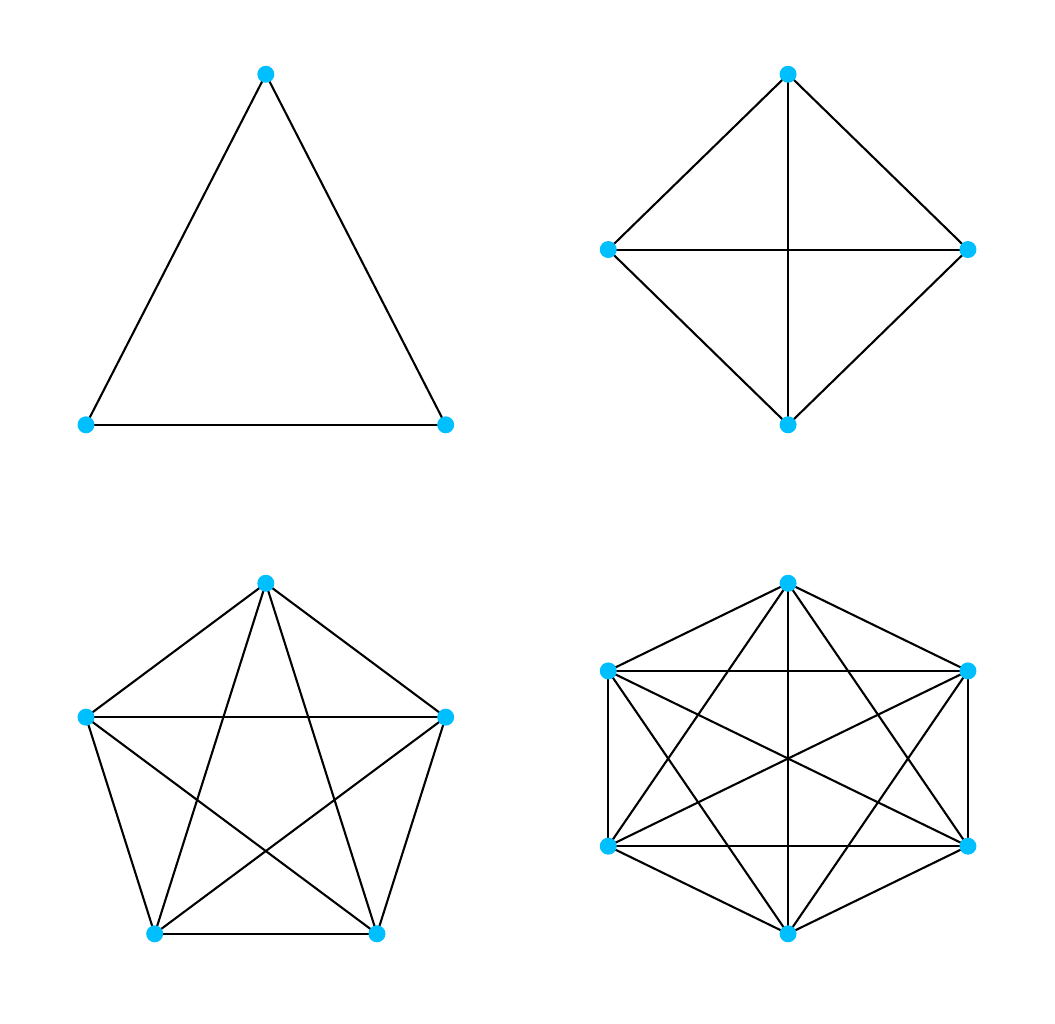}
	\caption{Examples of complete graphs $ K_{n} $ for low number of nodes $ n =3,4,5,6 $.}
	\label{fig:complete_graphs}
\end{figure}

Among search problems, classical algorithms often employ random walks  \cite{Random_Walk}, whereas quantum algorithms use quantum walks \cite{Ambainis_2014,QMS,Lemieux2020efficientquantum,Bezerra2023QuantumCO, search_on_bipartite, exp_speedup_childs}, enabling faster search and sampling methods. In this work, we utilize a specific type of quantum walk called the Szegedy quantum walk \cite{Szegedy_qw}. Instead of performing a traditional search on graphs, we exploit the properties of this quantum walk to test the completeness of a graph. An undirected graph is complete if all nodes are interconnected (see Fig.~\ref{fig:complete_graphs}).

We present an algorithm that distinguishes whether a graph is complete or not. It consists of two stages, involving Szegedy quantum walk and Quantum Phase Estimation (QPE) \cite{kitaev1995quantum,NielsenAndChuang}. In the first stage, we leverage a relationship between the total number of nodes $ n $ and an optimal number of marked nodes $ m^{\ast} $. This relationship is crucial for optimizing the running time and success probability of the Szegedy quantum walk search. For a complete graph  $\mathcal{G}_c$ (or $K_n$ in the traditional notation) with  $n$ nodes, specifying the number of marked nodes $ m^{\ast} $ ensures a bounded running time, which tends to a constant value of three time steps for large $ n$ and $m^{\ast}$.

The complexity of first stage is determined by this constant running time. The QPE subroutine in the second stage ensures a sufficient condition for completeness testing. QPE is widely used in various quantum algorithms, such as quantum counting and quantum search \cite{Bezerra2023QuantumCO,Quantum_counting_AA,NielsenAndChuang}. By knowing the number of marked nodes, we can predict the eigenvalues and eigenphases of the quantum walk evolution operator for a complete graph $\mathcal{G}_c$ with $n$ nodes. These eigenphases serve as indicators of the graph's completeness. By comparing the eigenphase, estimated via QPE, for the quantum walk evolution operator of the input graph $\mathcal{G}$ with the expected eigenphase of a correspondent complete graph $\mathcal{G}_c$, we can determine the completeness of $\mathcal{G}$.

The paper is organized as follows: in Section \ref{sec_II}, we begin with a brief overview of Markov chains and discrete-time quantum walks. We introduce the Szegedy quantum walk on bipartite graphs and define key quantities such as the hitting time and the probability distribution on marked nodes. Through analyzing these, we derive an optimality condition for the Szegedy quantum walk, revealing a linear relationship between the number of nodes in a complete graph and the number of marked nodes. The section concludes by examining the performance of the Szegedy quantum walk in terms of running time and success probability, utilizing this linear relationship.
Section \ref{sec_III} presents our algorithm for testing graph completeness. We provide a detailed description of the algorithm, derive the complexity, and compare it with other algorithms designed for similar purposes. This comparison underlines the quantum advantage offered by our completeness testing algorithm.
In Section \ref{sec_IV}, we discuss potential applications of the algorithm in graph structure analysis and in quantum-aided classical algorithms for network routing. 
Section \ref{sec_V} is devoted to conclusions. In addition, we include three appendices with the mathematical proofs and technical details supporting the results in the main text leading to the quantum algorithm for completeness testing.

\section{Discrete-Time Quantum Walks}\label{sec_II}

\subsection{Quantum Markov Chain}\label{sub_II_A}
A quantum walk is the quantum-mechanical analog of a classical random walk. In a discrete-time random walk, each step moves the walker to a new position based on a transition probability distribution. In the analogous quantum walk, the walker's position is represented by a quantum state, a superposition of multiple locations, and each step is described by a unitary operator. This operator depends on the underlying environment in which the walker moves. Quantum walks can also occur in continuous time, where the evolution is governed by a Hamiltonian \cite{Portugal_book}. This algorithm not only enhances the efficiency of completeness testing but also underscores the potential of quantum walks in solving fundamental problems in graph theory.
	
The Szegedy quantum walk \cite{Szegedy_qw,Portugal_book} is essentially the quantum analog of a classical Markov chain. A Markov chain is a discrete-time stochastic process, represented by a sequence of random variables, where each variable indicates a state within a state space $S$ and the future state depends only on the present state.  Denote the graph as $\mathcal{G}=\{V,E\}$,  with $ V $ and $ E $ their sets of vertices and edges,  respectively.  The transition probabilities from one state to another, denoted as  $p_{ij}$,  are time-independent, forming a transition matrix $P$. Every Markov chain with such transition probabilities can be represented by a directed graph (digraph) if $p_{ij}\neq p_{ji}$, or a undirected graph if $P$ is symmetric. In this representation, the vertex set $V=\{1, \dots, i, \dots, n\}$ corresponds to the state space $S=\{ s_1, \dots, s_n \}$, and the edge set $E$ contains an arc $(i,j)$ if and only if $p_{ij}>0$.

\begin{figure}[t]
	\centering
  \includegraphics[width=0.5\textwidth]{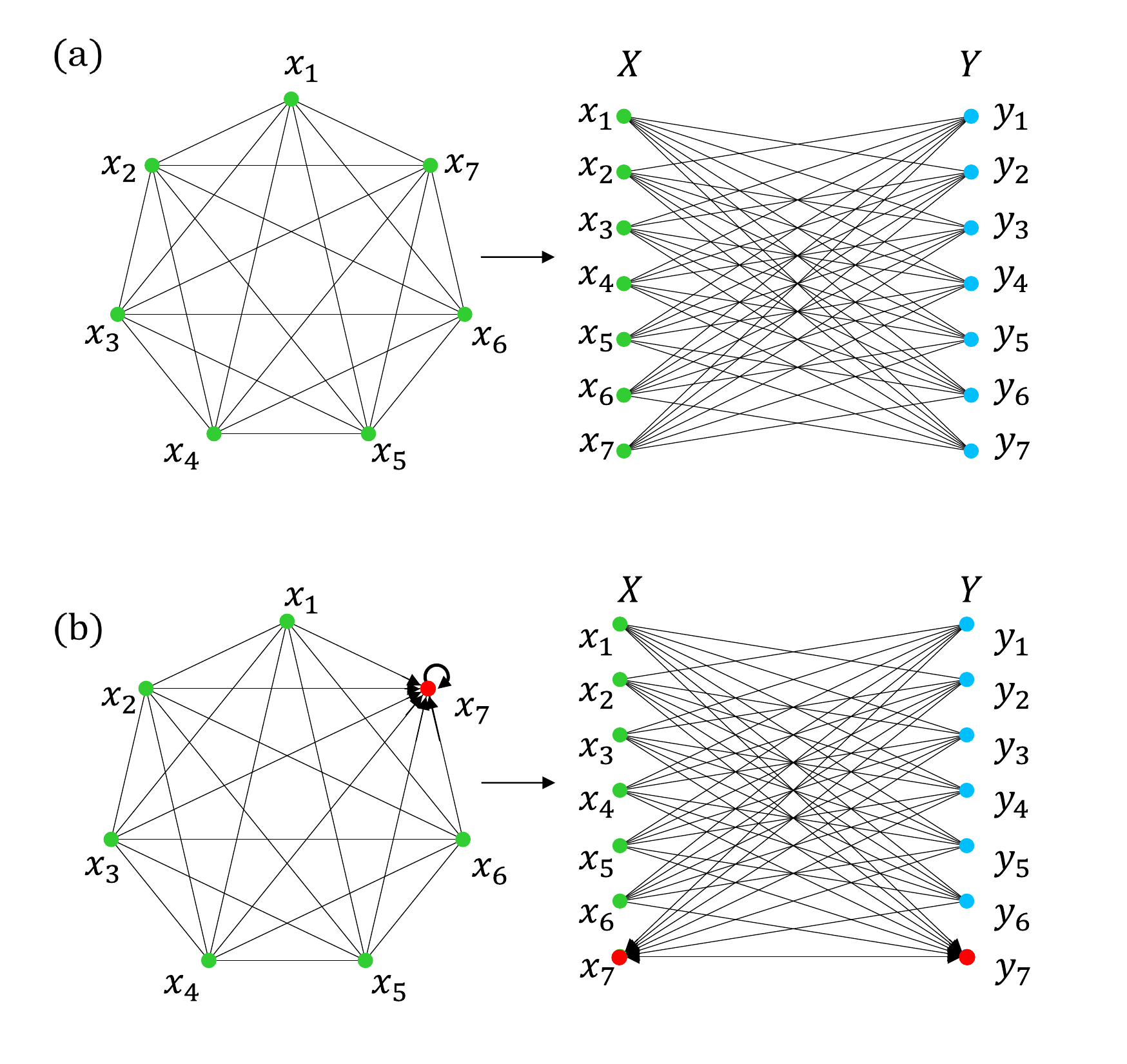}
  \caption{(a) Original unmarked graph $\mathcal{G}$ and its duplicated graph for the Szegedy quantum walk. To each edge $(i,j)$ of the underlying graph correspond two edges of the bipartite one $(x_i, y_j)$ and $(y_i, x_j)$. (b) Similarly with a marked node in red.  The marked vertices are absorbing states $s_i$ with only incoming directed links and no outgoing ones. In the duplicated graph the edges connecting the twin marked edges are added.}
  \label{fig:Szegedy_complete_graphs}
\end{figure}

Using the formalism of a Markov chain, we can describe a random walk on this type of graph. Similarly, the Szegedy model is a discrete-time, coinless quantum walk on symmetric bipartite digraphs. A bipartite digraph can be derived from a simple graph through a duplication process \cite{Portugal_book}, facilitating the application of the Szegedy quantum walk (see Fig.~\ref{fig:Szegedy_complete_graphs}).	
	
When implementing a search algorithm on a graph, whether quantum or classical, the process involves marking the searched vertices, which correspond to searched states in the state space $S$.  Additionally, a final time for the walk evolution must be defined. The marked vertices have only incoming directed links and no outgoing ones, corresponding to absorbing states $s_i$ where $p_{ii}=1$ and $p_{ij}=0$ for all $j\neq i$.	
	
	\subsection{Szegedy Quantum Walk}\label{sub_II_B}
	
	The quantum walk on a graph $\mathcal{G}$ with vertices $V$,  is associated with the Hilbert space $\mathcal{H}^n$ where $n$ is the number of vertices ($n=|V|$). Its computational basis is spanned by states associated with the vertices $x$, $\{\ket{x}: x \in V\}$. A bipartite graph is a graph whose vertices can be divided into two disjoint and independent sets such that every edge connects a vertex in the first set to one in the second set and vice versa.

The Szegedy quantum walk we introduce takes place on a bipartite graph derived from an underlying directed graph through a duplication process. For a bipartite graph, the associated Hilbert space is 
$\mathcal{H}^{n}\otimes \mathcal{H}^{n}=\mathcal{H}^{n^2}$.
After the duplication process, if we call the two groups of vertices $X$ and $Y$, with the corresponding computational bases $\{\ket{x}: x \in X\}$ and $\{\ket{y}: y \in Y\}$, the computational basis for the bipartite quantum walk will be $\{\ket{x,y}:x\in X,y\in Y\}$, with $\mathcal{H}^{n^2}=\text{span}\left\{\ket{x,y}:x\in X,y\in Y \right\}$. We define the transition probabilities $p_{xy}$ as the inverse of the degree of vertex $x$ if $x$ is adjacent to $y$, otherwise $p_{xy}=0$, and similarly $q_{yx}$ as the inverse of the degree of vertex $y$ if $y$ and $x$ are adjacent, otherwise $q_{yx}=0$. We define $p_{xy}$ and $q_{yx}$ as elements respectively of matrix $P$ and $Q$.

Given a transition matrix $P$ associated with the graph $\mathcal{G}$, the evolution operator of the quantum walk on the bipartite graph is given by:
	\begin{gather}\label{eq_evolution_operator}
		W_P=\mathcal{R}_B \mathcal{R}_A,
	\end{gather}
where $ \mathcal{R}_A $ and $ \mathcal{R}_B $
are reflection operators related to the subspaces \(\mathcal{A}\) and \(\mathcal{B}\) of the Hilbert space \(\mathcal{H}^{n^2}\). These subspaces correspond to the two registers \(\ket{x}\) and \(\ket{y}\) of the bipartite graph after the duplication process.

Through \( W_P \), the superposition of nodes in the initial state of the walker \( \ket{\psi(0)} \) evolves. The reflections are defined as follows:
	\begin{gather}\label{eq_reflection_operators}
 \mathcal{R}_A = 2\Pi_A - I_{n^2}, \quad \mathcal{R}_B = 2\Pi_B - I_{n^2} ,
 	\end{gather}
where:
	\begin{gather}\label{eq_projection_operators}
\Pi_A = \sum_{x \in X} \ket{\alpha_x}\bra{\alpha_x}, \quad \Pi_B = \sum_{y \in Y} \ket{\beta_y}\bra{\beta_y}, 
	\end{gather}
are the projections onto the subspaces \(\mathcal{A} = \text{span}\{\ket{\alpha_x} : x \in X\}\) and \(\mathcal{B} = \text{span}\{\ket{\beta_y} : y \in Y\}\).

Each vector \(\ket{\alpha_x}\) represents a vertex \( x \) in the first set \( X \) of the bipartite graph and its weighted connections with the vertices \( y \) in the second set \( Y \) via the transition matrix elements \( p_{xy} \). Similarly, each vector \(\ket{\beta_y}\) represents a vertex \( y \) in \( Y \) and its connections with vertices \( x \) in \( X \), replacing the matrix $P$ with the matrix $Q$. These vectors are defined as:
	\begin{gather}\label{eq_alphax_and_betay}
 \ket{\alpha_x} = \ket{x} \otimes \sum_{y \in Y} \sqrt{p_{xy}} \ket{y}, \quad \ket{\beta_y} = \sum_{x \in X} \sqrt{q_{yx}} \ket{x} \otimes \ket{y}.
	\end{gather}

In the case of a bipartite graph derived from a duplication procedure of a simple graph, the transition probabilities are symmetric, \( p_{xy} = q_{yx} \). This ensures the correct representation of the bipartite graph's structure within the Szegedy quantum walk framework.

In the case where we mark randomly some of the vertices in the graph $\mathcal{G}$ and perform a quantum walk of this kind, we are essentially conducting a quantum search on the graph \cite{Grover,NielsenAndChuang,Portugal_book} through the Szegedy quantum walk. Defining the subset of marked vertices as $M=\{x_1,\dots,x_m\}$, we use a new transition matrix \(P'\), defined as:
\begin{gather}\label{system_transition_matrix}
	p'_{xy}=
	\begin{cases}
		p_{xy} \quad x \notin M \\
		\delta_{xy} \quad x \in M			
	\end{cases}.
\end{gather}
Thus, the Szegedy quantum walk on the bipartite graph takes two reflections which periodically bring the initial state \(\ket{\psi(0)}\) closer to the marked vertex states \cite{Portugal_book}. 

Once the evolution operator is defined and the initial state \(\ket{\psi(0)}\) is set, the final state after \(t\) steps is given by:
\begin{gather}\label{eq_equation_evolution}
	\ket{\psi(t)}=W_{P'}^t\ket{\psi(0)}.
\end{gather}
The probability of finding the walker on a marked node exhibits oscillatory behavior \cite{Portugal_book,Szegedy_qw}, hence it is necessary to define a finite running time for our search algorithm.

	\subsection{Hitting Time and Probability Distribution on the Marked Nodes}\label{sub_II_C}

As we anticipated, we need to establish how many steps of the evolution to run in order to obtain a high probability of finding the walker on the marked nodes. To this end, we introduce an averaged distance
between the intitial and final states after the walk evolution:
\begin{defn}{Time Averaged Distance (TAD).}
\begin{gather}\label{eq_definition_F(T)}
	F(T):=\frac{1}{T+1}\sum_{t=0}^{T}\bigl\|\ket{\psi(t)}-\ket{\psi(0)}\bigr\|^2,
\end{gather}
\end{defn}
where $\ket{\psi(t)}$ is defined by Eq.(\ref{eq_equation_evolution}). This function can be understood as an average distance between the initial state and the evolved state at the final time $T$.
Then,  for the Szegedy quantum walk on the bipartite digraph,  the quantum hitting time, denoted as $t_{\text{h}}$ is defined as the minimum time at which the following inequality holds \cite{Portugal_book,Szegedy_qw}:
\begin{defn}{Hitting Time.}
\begin{gather}\label{eq_definition_F(T)th}
	F(t_{\text{h}})\geq 1-\frac{m}{n}.
\end{gather}
\end{defn}
The value on the right is the difference between the uniform probability distribution on all the nodes and the uniform probability distribution on the marked nodes.

This definition of the hitting time is a generalization of the classical hitting time definition for a random walk, that is, the expected time for a classical walker to reach a marked vertex for the first time, once given the initial conditions. The probability of finding the walker on the marked vertices, which we are trying to maximize, is calculated with the projector $\mathcal{P}_M$:
\begin{gather}\label{eq_definition_probability_projector}
\mathcal{P}_M:=	\sum_{x \in M}\sum_{y \in Y}\ket{x,y}\bra{x,y}.
\end{gather}
At time $t$, it is given by:
\begin{gather}\label{eq_definition_probability}
	P_M(t):=\bra{\psi(t)}\mathcal{P}_M\ket{\psi(t)}.
\end{gather}
Let us call $t_{max}$ the time associated with a maximum of the probability $P_{M}(t)$. In order to set a running time $t_{h}$ it is necessary to set the initial state $\ket{\psi(0)}$ and calculate the spectral decomposition of $W_{P'}$. The initial state for the Szegedy's quantum walk is defined using the transition matrix of the underlying graph $\mathcal{G}$ with no marked vertices:
\begin{gather}\label{eq_initial_state_general}
	\ket{\psi(0)}=\frac{1}{\sqrt{n}}\sum_{x\in X, y \in Y}\sqrt{p_{xy}}\ket{x,y}.
\end{gather}
We focus our analysis on the case of a complete graph with $n$ nodes. We denote the transition matrices without and with $m$ marked vertices as $P_c$ and $P'_c$, respectively. The matrix $P'_c$ is defined as:
\begin{gather}\label{system_transition_matrix_complete_graph}
	p_{_c xy}'=
	\begin{cases}
		\frac{1-\delta_{xy}}{n-1}, & x \notin M; \\
		\delta_{xy}, & x \in M.
	\end{cases}
\end{gather}
For a complete graph with $n$ nodes, the initial state given in Eq.(\ref{eq_initial_state_general}) simplifies to:
\begin{gather}\label{eq_initial_state_complete_case}
	\ket{\psi(0)} = \frac{1}{\sqrt{n(n-1)}} \sum_{x,y=1}^n (1-\delta_{xy}) \ket{x,y}.
\end{gather}

In this complete graph scenario, the spectral decomposition of $W_{P'_c}$ can be calculated analytically \cite{Portugal_book, Szegedy_qw}. This also holds true for the hitting time $t_{h}$, the probability function $P_M(t)$, the maximum time $t_{max}$, and the function $F(T)$. For a complete graph, only specific eigenspaces of $W_{P'_c}$ are involved in the dynamics. When the initial state in Eq.(\ref{eq_initial_state_complete_case}) is expressed in terms of the eigenbasis of $W_{P'_c}$, it only involves a subset of the eigenvectors. The set $\sigma(W_{P'_c})$ of eigenvalues and corresponding eigenvectors of $W_{P'_c}$ \cite{Portugal_book} is:
\begin{gather}\notag
	\sigma(W_{P'_c}) = \left\{ e^{\pm2i\theta_1}, \left\{\ket{\theta^{+}_j} \ket{\theta^{-}_j}\right\}_{j \in \left[1, n-m-1\right]} \right.; \\ \notag
	\quad \! e^{\pm2i\theta_2}, \left\{\ket{\theta^{+}_2}, \ket{\theta^{-}_2}\right\}; \\ \label{eq_eigenvalues_eigenvectors_complete_case}
	\left. \qquad \quad \: 1, \left\{\ket{\theta_0}, \ket{\theta_j}\right\}_{j \in \left[n-m+1, n\right]}\right\}
\end{gather}
The eigenphases associated with the eigenvectors of $W_{P'_c}$, denoted as $\Theta=\left\{\theta_0, \theta_1, \theta_2\right\}$, are defined for the complete case as follows:
\begin{gather}\label{eq_eigenphases_complete_case}
	\theta_0 = 0, \quad \cos(\theta_1) = \frac{1}{n-1}, \quad \cos(\theta_2) = \frac{n-m-1}{n-1}.
\end{gather}
Among these, only the eigenvectors associated with the eigenvalues connected to $\theta_2$ and $\theta_0$ are involved in the dynamics.  Specifically, the initial state can be decomposed as:
\begin{gather}\label{eq_initial_state_complete_case_decomposition_eigenvectors_W}
	\ket{\psi(0)} = c^{+}\ket{\theta_2^{+}} + c^{-}\ket{\theta_2^{-}} + \ket{\theta_0},
\end{gather}
with inital coefficients $ c^{\pm} $, implying that the time-evolved state $\ket{\psi(t)}$ also involves only these eigenvectors:
\begin{gather}\label{eq_final_state_complete_case_decomposition_eigenvectors_W}
	\ket{\psi(t)} = c^{+} e^{i2\theta_2 t} \ket{\theta_2^{+}} + c^{-} e^{-i2\theta_2 t} \ket{\theta_2^{-}} + \ket{\theta_0}.
\end{gather}

It is important to note that the phases are within the interval $\left[0, \pi/2\right]$ by definition \cite{Portugal_book}. To explicitly indicate the dependence on $n$ and $m$ for the quantities $P_M(t)$, $F(T)$, and $t_{max}$ in the context of a complete graph, we denote them as $P_M(t,n,m)$, $F(t,n,m)$, and $t_{max}(n,m)$, respectively.

\subsection{Optimality Condition for the Szegedy Quantum Walk}\label{sub_II_D}

It is proven that the hitting time $t_{h}$ defined by Eq.(\ref{eq_definition_F(T)th}) provides a good approximation of $t_{max}(n,m)$ \cite{Portugal_book}, even if they do not coincide precisely. Furthermore, the quantum hitting time on a finite bipartite graph is quadratically smaller than the classical hitting time of a random walk on the underlying graph \cite{Szegedy_qw,Szegedy2004spectra, Portugal_book}. Thus, using the Szegedy quantum walk search provides a relative quantum advantage.

What happens if we require the hitting time to coincide with one of the maxima of $P_M(t,n,m)$? In other words, if we want to achieve the best performance both in terms of running time and probability, we wonder about the requirements on the other variables, such as the graph structure. We restrict our study to the complete case and use the known analytical expressions of $t_{max}(n,m)$ and $F(t,n,m)$.

The function $F(t,n,m)$ for the complete case \cite{Portugal_book} is given by:
\begin{gather}\label{eq_F(T)_complete_case}
    F(t,n,m) = \frac{2(n-1)(n-m)\left(2t+1-U_{2t}\left[\frac{n-m-1}{n-1}\right]\right)}{n(2n-m-2)(t+1)},
\end{gather}
where $U_{2t}$ is a Chebyshev polynomial of the second kind. The expression for $t_{max}(n,m)$ is:
\begin{gather}\label{eq_t_max_analytical}
    t_{max}(n,m) = \frac{\arctan\left(\frac{\sqrt{2n-m-2}}{\sqrt{m}}\right)}{2\arccos\left(\frac{n-m-1}{n-1}\right)}.
\end{gather}

For completeness, we report the expression for the probability $P_M(t,n,m)$ for the complete graph case:
	\begin{gather}\notag 
		P_{M}(t,n,m)=\frac{m(n-m)}{n(n-1)}\left(\frac{n-1}{2n-m-2}T_{2t}\left[\frac{n-m-1}{n-1}\right] \right.+ \\
		\left.+ U_{2t-1}\left[\frac{n-m-1}{n-1}\right]+\frac{n-m-1}{2n-m-2}\right)^2+\frac{m(m-1)}{n(n-1)},\label{eq_probability_complete_graph}
	\end{gather}
where $U_{2t-1}$ and $T_{2t}$ are Chebyshev polynomials of the second and first kinds, respectively.

To require the coincidence of the hitting time with the maximum of $P_M(t,n,m)$, we solve the system of equations composed of Eqs.(\ref{eq_definition_F(T)th}) (when equality holds) and (\ref{eq_t_max_analytical}).

In this analysis, we have three variables: $T$, $n$, and $m$, but only two equations. By treating the number of marked nodes $m$ as a parameter, we can rephrase our question: How many nodes should we mark randomly in a complete graph with $n$ nodes to make the hitting time coincide with at least one $t_{max}$?

To find the answer, we substitute the variable $T$ in Eq.(\ref{eq_definition_F(T)th}) with the expression for $t_{max}(n,m)$ from Eq. (\ref{eq_t_max_analytical}). This substitution yields an equation that describes the relationship between $n$ and $m$ that meets our requirement:
\begin{defn}{Optimality Condition (OC).}
\begin{gather}\notag
    F\left( t_{max}(n,m), n, m \right) = 1 - \frac{m}{n}, \\
    t_{max}(n,m) = \frac{\arctan\left(\frac{\sqrt{2n-m-2}}{\sqrt{m}}\right)}{2\arccos\left(\frac{n-m-1}{n-1}\right)}.
    \label{system_F(T)_tmax}
\end{gather}
\end{defn}

Solving this system reveals a linear relationship between $n$ and $m$. 
\begin{prop}{Solution of the Optimality Condition.}\label{solution_of_the_optimality_condition}
	\begin{gather}\label{eq_linear_relationship_numerical}
		n=a m+1.
	\end{gather}
	The constant $ a $ is the solution of the following transcendental equation:
		\begin{gather}\label{solution_of_the_optimality_condition_eq}
		\Scale[0.9]{\left(4 \sqrt{2} a^{3/2}+2 a-3\right) \arctan\left(\sqrt{2 a-1}\right)-2 \pi  \left(\sqrt{2} a^{3/2}-1\right)=0},
	\end{gather}
	whose numerical solution yields $ a = 1.44512  $.  \qed
\end{prop}

We refer to Appendix \ref{Appendix_A} for an analytical solution of the optimality condition. 
This relationship can be interpreted as a method for calculating the \textit{optimal} number of nodes $m^*$ to mark in the graph. From Eq.(\ref{eq_linear_relationship_numerical}), we can calculate the optimal $m^*$ corresponding to a fixed $n$. To ensure a meaningful value in terms of the graph's nodes, we need to round $m^*$ to its closest integer. By ensuring that the hitting time $t_{h}$ coincides with the maximum of the probability $P_M(t,n,m)$, we maximize the probability of being on a marked node in the shortest possible time.

\subsection{Boundaries for Time and Probability}\label{sub_II_E}
The linear relationship between \( n \) and \( m \) found in the previous section guarantees the coincidence of the hitting time \( t_{h} \) with the first maximum point \( t_{max}(n,m) \) of the probability (Eq.(\ref{eq_t_max_analytical})). This linear relationship influences both the behavior of \( t_{max} \) itself and the probability value at this time for growing \( n \). Let us denote the \( t_{max} \), after introducing the linear relationship (\(\ref{eq_linear_relationship_numerical}\)) as \( t^*_{max} \). Using the ansatz described in Appendix \ref{Appendix_A}, \( m=\frac{n-1}{a} \), the dependence on \( n \) disappears. Thus, from Eq.(\ref{eq_t_max_analytical}), we obtain the following expression for \( t^*_{max} \):

\begin{gather}\label{eq_t*max}
t^*_{max}= \frac{\arctan\left(\sqrt{2a-1}\right)}{2 \arccos\left(\frac{a-1}{a}\right)}.
\end{gather}

Thus, \( t^*_{max} \) is a constant value, and with \( a \approx 1.4451 \) it becomes \( t^*_{max} \approx 0.3745 \). This guarantees that the hitting time remains finite independently of \( n \) when the number of marked nodes is chosen through the linear relationship in Eq.(\ref{eq_linear_relationship_numerical}). This is not generally true for complete graphs if a different \( m \) is chosen; the asymptotic value of the hitting time and the \( t_{max}(n,m) \), for \( n \gg m \) scales as \( \propto \sqrt{\frac{n}{2m}} \).

\begin{figure}[t]
		\centering
		\includegraphics[width=1.1\columnwidth]{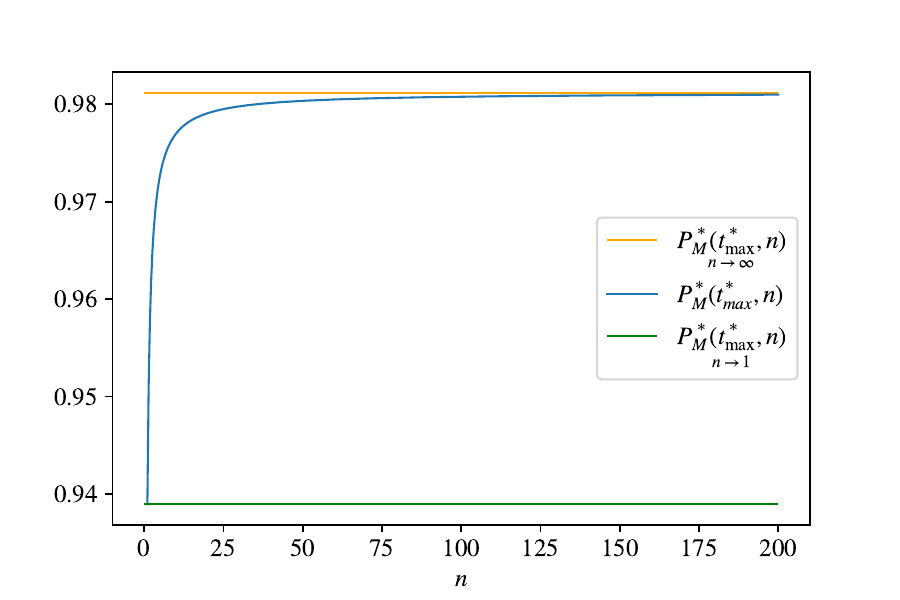}
		\caption{The probability of finding the walker on a marked node reaches its maximum value at time $t^*_{\text{max}}$. Introducing the linear relationship presented in Eq.(\ref{eq_linear_relationship_numerical}) confines this maximum probability value, denoted as $P^*_{M}(t^*_{\text{max}},n)$, between the numerical limits $\lim_{n\to 1}P^*_{M}(t^*_{\text{max}},n)$ and $\lim_{n\to \infty}P^*_{M}(t^*_{\text{max}},n)$. In the plotted graph, the trend of $P^*_{M}(t^*_{\text{max}},n)$ is illustrated for $n$ values ranging from $1$ to $200$, along with its limits as $n$ approaches $1$ and $\infty$. The numerical values for these boundaries are provided in Eqs.(\ref{eq_probability_boundaries_numerical_1}) and (\ref{eq_probability_boundaries_numerical_2}), respectively.		}
		\label{fig_P_t*max_boundaries}
	\end{figure}

We should note that the Szegedy quantum walk we are analyzing is a discrete-time quantum walk and thus requires a discrete number of steps. The closest integer to the computed \( t^*_{max} \) is, however, zero, making it unsuitable as a running time for a discrete quantum walk. Since the probability \( P_{M}(t,n,m) \) is a periodic function with a period \( \mathcal{T} \) equal to \( \pi/\arccos\left(\frac{-m+n-1}{n-1}\right) \), we opt to select the second maximum of the probability as our running time. The second maximum also possesses a constant value when introducing the linear relationship (\(\ref{eq_linear_relationship_numerical}\)). The periodic probability behaves consistently as described in Fig.\ref{fig_P_t*max_boundaries} for its second maximum.

Upon introducing the linear relationship (\(\ref{eq_linear_relationship_numerical}\)) into the period \( \mathcal{T} \), the value of the second maximum is:

\begin{gather}\label{eq_tmax_second}
t^*_{max}+\frac{\pi }{\arccos\left(\frac{a-1}{a}\right)}\approx  2.8724.
\end{gather}

The closest integer to this limit is \( 3 \). Let us call this value \( t^* \); this will be the running time for the first part of our algorithm (see Sec.\ref{sub_III_A}).

We observe peculiar behavior also for the probability \( P_M(t,n,m) \). When we introduce the linear relationship and evaluate the probability at the first or second maximum, it exhibits an upper bound for \( n \to \infty \) and a lower bound for \( n \to 1 \) (see Fig.\ref{fig_P_t*max_boundaries}).

Let us call \( P^*_M(t^*_{max},n) \) the probability \( P_M(t,n,m) \) when \( t=t^*_{max} \) and when the linear relationship (Eq.(\ref{eq_linear_relationship_numerical})) is introduced through the ansatz \( m=\frac{n-1}{a} \). The limits of \( P_M(t^*_{max},n) \) for \( n \to 1 \) and for \( n \to \infty \) are, respectively:

\begin{subequations}
\begin{align}
\lim_{n\to 1}P^*_{M}(t^*_{max},n)\approx 0.9389, \label{eq_probability_boundaries_numerical_1} \\
\lim_{n\to \infty}P^*_{M}(t^*_{max},n) \approx 0.9812. \label{eq_probability_boundaries_numerical_2}
\end{align}
\end{subequations}
These values are plotted together with \( P^*_{M}(t^*_{max},n) \) in Fig.\ref{fig_P_t*max_boundaries}.
We observe that the probability remains consistently close to $1$, indicating that it is highly improbable to find the walker on one of the unmarked vertices when measuring the position at time $t^*_{\text{max}}$ with $m^*$ nodes marked. It's noteworthy that these boundaries also apply to the second maximum in Eq.(\ref{eq_tmax_second}). 
Consequently, as $P^*_M(t^*_{\text{max}},n)\approx1$ for a complete graph, evolving the initial state through:
\begin{gather}\label{eq_final_state_t*max}
    \ket{\psi(t^*_{\text{max}})}=W_{P'_c}^{t^*_{\text{max}}}\ket{\psi(0)},
\end{gather}
and measuring the position $\ket{x}$, implies $x \in M$ if the graph is complete. In other words: $\mathcal{G}$ is complete $\Rightarrow P^*_M(t^*_{\text{max}},n)\approx1 \Rightarrow x \in M$. Since if $A\Rightarrow B$ is true, then $\neg B \Rightarrow \neg A$ is also true, we can infer the following, based on the measurement of the walker position after the quantum walk evolution:

\begin{itemize}
    \item Not finding a marked node $\Rightarrow$ the graph is incomplete.
\end{itemize}
However, it's important to note that:
\begin{itemize}
    \item Finding a marked node $\nRightarrow$ the graph is complete,
\end{itemize}
since the probability of locating the walker on a marked node is not strictly zero, even if the graph is incomplete. These observations will be instrumental in section \ref{sub_III_A} for constructing the core of the completeness testing algorithm.

\subsection{Quantum Phase Estimation and Spectral Radius}\label{sub_II_F}

In the algorithm detailed in Sec.\ref{sec_III}, the quantum phase estimation (QPE) \cite{NielsenAndChuang,kitaev1995quantum} plays a fundamental role. Below, we present the key result used in the algorithm, providing the theoretical foundation required for understanding its application.

Provided that $\ket{\theta_2^{+}}$ is an eigenstate of the evolution operator $W_{P_c'}$ of a complete graph with $n$ nodes and $m$ marked ones, according to Eq.~(\ref{eq_eigenvalues_eigenvectors_complete_case}), we can state the following proposition:
\begin{prop}\label{proposition_2}
    \textbf{Graph Completeness Indicator.}\\
    The QPE, when provided with the input state $\ket{0}^{\otimes p}\ket{\theta_2^{+}}$ and the unitary operator $W_{P'}$, estimates the eigenphase $\theta_2 = \arccos{\frac{n-m-1}{n-1}}$ if and only if $P'$ is the transition matrix of a complete graph with \( n \) nodes and \( m \) marked nodes. In other words, this holds if and only if \( P' = P_c' \).
\end{prop}

The eigenphase $\theta_2$ is related to the spectral radius of the adjacency matrix of the complete graph with $n-m$ nodes \cite{stanley1987bound}. The spectral radius of the adjacency matrix of an incomplete graph with $n-m$ nodes is always smaller than that of the complete graph \cite{Bound_spectral_radius}. Therefore, the eigenphase $\theta_2$ is a reliable indicator of the completeness of the graph. This property is crucial for the completeness testing algorithm in Sec.\ref{sec_III}. 

The QPE takes as inputs a unitary operator $U$ and a state $\ket{0}^{\otimes p}\ket{u}$, encoded in two registers. The first register consists of $p$ qubits and determines the precision of the phase estimation, while the second register encodes the state $\ket{u}$. If $\ket{u}$ is an eigenstate of $U$ with eigenvalue $e^{i2\pi \theta_u}$, the output is an estimate of $\theta_u$ with $p$ digits of precision. If the input state $\ket{\psi}$ in the second register is a linear combination of the eigenstates of $U$, the eigenphase $\theta_u$ is estimated with a probability given by $|c_u|^2=|\braket{u|\psi}|^2$.

\begin{proof}
    When the input state $\ket{0}^{\otimes p}\ket{\theta_2^{+}}$ and the operator $W_{P'_c}$ (the evolution operator of the Szegedy quantum walk on a complete graph) are provided to the QPE subroutine, the phase $\theta_2/\pi$ is estimated with \( p \) digits of precision, obtaining $\theta_2$ with the same precision. The mechanism of the QPE ensures that the consequent in Proposition \ref{proposition_2} implies the antecedent: specifically, if \( P' = P'_c \), then QPE will estimate $\theta_2$ when the input state and operator are $\ket{0}^{\otimes p}\ket{\theta_2^+}$ and $W_{P'_c}$, respectively.

    Conversely, if the initial state $\ket{0}^{\otimes p}\ket{\theta_2^{+}}$ is used with an operator $W_{P'}$ related to an incomplete graph, QPE will estimate a phase $\theta_j$, which typically differs from $\theta_2$. This distinction helps in identifying whether the graph is complete.

    To ensure that this method does not mistakenly estimate the phase $\theta_2$ for an incomplete graph by chance, we recognize that the eigenvalues $e^{\pm 2i\theta_2}$ are derived from the maximum eigenvalue of the adjacency matrix of a complete graph $\mathcal{G}_c$ with \( m \) marked nodes \cite{Portugal_book, Bound_spectral_radius}. This relationship supports that the antecedent (having a complete graph) in Proposition \ref{proposition_2} indeed implies the consequent (estimating $\theta_2$ via QPE).

\end{proof}

For a complete graph it can be shown \cite{Portugal_book} that the eigenvalues of the evolution operator $W_{P'_c}$ that differ from $+1$ can be obtained from the transition matrix of the underlying graph with no marked nodes $P_c$. More specifically, an auxiliary matrix $C$ facilitates finding the spectrum of $W_{P'_c}$, with $c_{xy}=\sqrt{p_{xy}q_{yx}}$, which can be written in the following form when the graph is complete and the last $m$ nodes marked: 
\begin{gather}\label{C_matrix_complete_graph}
	C = \left[\begin{matrix}
			P_{_c n-m} & 0\\
			0     & I_m
		\end{matrix}\right].
\end{gather}
Where $P_{_c n-m}$ is the matrix obtained from $P_c$ eliminating $m$ rows and columns corresponding to the marked vertices. From the submatrix $P_{_c n-m}$, we can derive the eigenvalues of $W_{P'_c}$ corresponding to $e^{\pm2 i\theta_2}$. For a complete graph, the characteristic polynomial of $P_{_c n-m}$ is:
\begin{gather}\label{characteristic_polynomial_complete_case}\Scale[0.9]{
		\text{det}(P_{_c n-m}-\lambda I_{n-m})=\left(\lambda- \frac{n-m-1}{n-1}\right)\left(\lambda+\frac{1}{n-1}\right)^{n-m-1}.}
\end{gather}
The eigenvalue $\lambda = \frac{n-m-1}{n-1}$ is the largest eigenvalue of $P_{_c n-m}$ and relates to $\theta_2$ by $\cos(\theta_2) = \frac{n-m-1}{n-1}$. Naming $A_{_c n-m}$ the adjacency matrix corresponding to the same graph, this eigenvalue corresponds to the spectral radius $\rho(A_{_c n-m})$, i.e., $\rho(A_{_c n-m}) = n-m-1$. 
According to a theorem in \cite{Bound_spectral_radius} the spectral radius of the adjacency matrix of a generic graph $\mathcal{G}$ with $n$ nodes and $|E|$ edges is bounded by:
\begin{gather}\label{matrix_radius_inequality}
	\rho(A) \leq \sqrt{2|E| - n + 1},
\end{gather}
where equality holds for the complete graph and the star graph \cite{Bound_spectral_radius, brualdi1985spectral, stanley1987bound}. Hence, for a graph with $n$ nodes and $m$ marked ones that is not complete, the following inequalities hold:
\begin{gather}\notag
	\rho(A_{n-m}) \leq \sqrt{2|E_{n-m}| - (n-m) + 1} <\\ \label{eq_chain_inequal_spectral_radius}
	  < \sqrt{2|E_{_c n-m}| - (n-m) + 1} = \rho(A_{_c n-m}),
\end{gather}
where $\rho(A_{n-m})$ is the spectral radius of the adjacency matrix of an incomplete graph with $n$ nodes, after removing $m$ rows and columns corresponding to $m$ marked nodes, which can be easily seen it is $\rho(A_{_c n-m})=n-m-1$. Notice that $|E| < |E_c|$ by definition. From these inequalities, we conclude that $\rho(A_{n-m}) < \rho(A_{_c n-m})$. Based on the analysis of the spectral radius of nonnegative matrices \cite{nn_matrices}, and supported by our simulations in \ref{Appendix_C}, we assume this inequality holds also for the corresponding matrices $P_{n-m}$ and $P_{_c n-m}$. This ensures that an incomplete graph with $n$ nodes and $m$ marked ones will always exhibit a smaller spectral radius compared to a complete graph, and thus it would never have $e^{\pm 2i\theta_2}$ as eigenvalues.

Therefore, finding the eigenphase $\theta_2$ using the QPE indicates the completeness of the graph. Note that for a fixed $n$, the closer $|E|$ is to $|E_c|$, the closer $\rho(A_{n-m})$ is to $\rho(A_{_c n-m})$. Thus, the "worst case" scenario, where the eigenvalues of an incomplete graph are closest to those of a complete graph, occurs when only one edge is missing, i.e., $|E| = |E_c| - 1$.

\section{Quantum Algorithm}\label{sec_III}
	
	\subsection{Completeness Testing Agorithm}\label{sub_III_A}

We present an algorithm designed to test the completeness of a given undirected graph $\mathcal{G}=\left( V,E \right)$, where $V=\{1,\dots,n\}$ is the set of vertices and $E$ is the set of edges. The graph has $n$ nodes (or vertices) and is represented by its transition matrix $P$. We introduce the position operator $X$ and its associated eigenbasis, comprising the positions of the vertices $V$ of the graph: $\{1,\dots, n\} \rightarrow \{\ket{1},\dots,\ket{n}\}$, usually denoted $\ket{x}$.
\begin{figure}[t]
	\centering
		\includegraphics[width=0.7\textwidth]{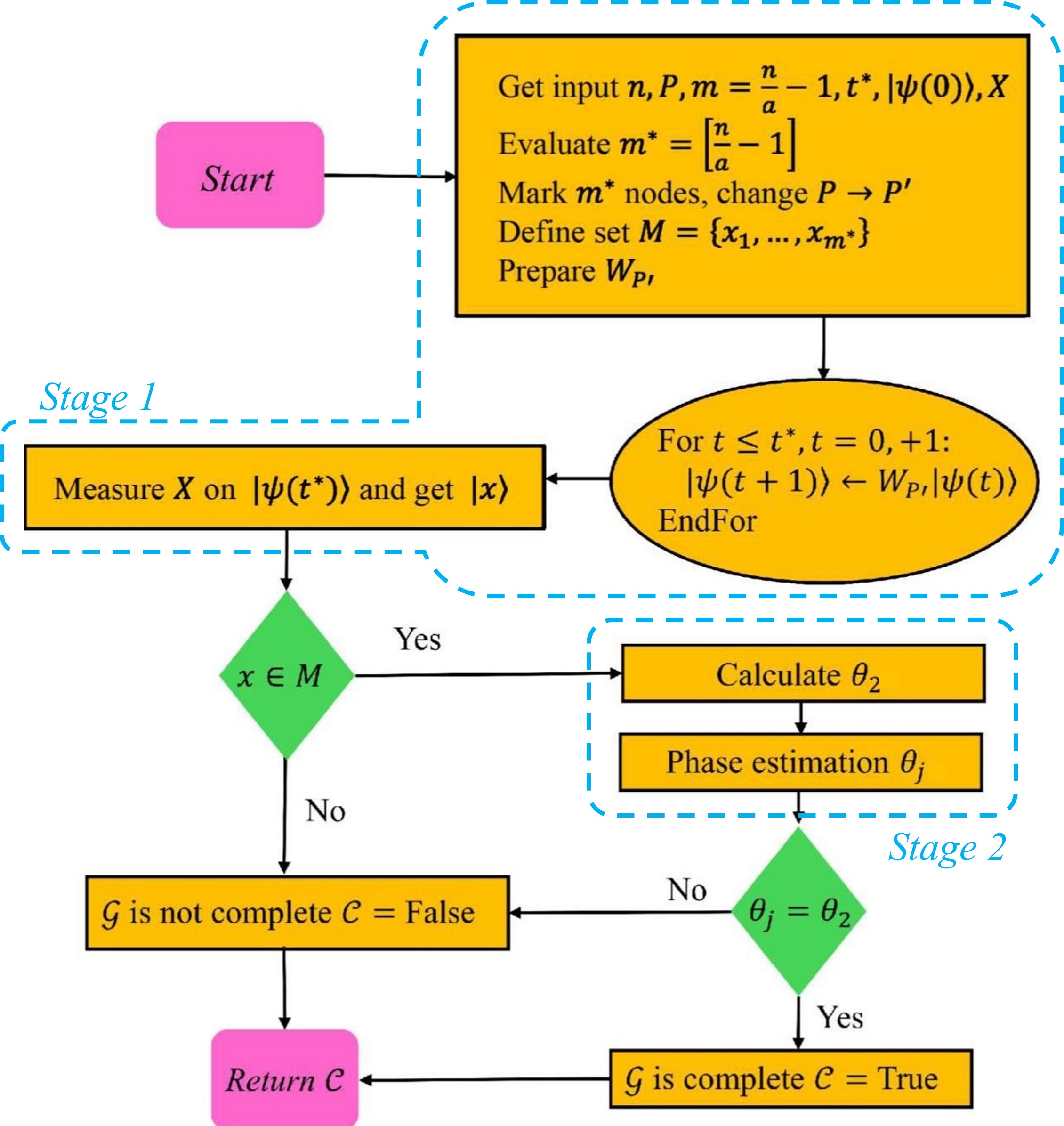}
		\caption{Flowchart of the completeness testing quantum algorithm as described in Subsection \ref{sub_III_A}. Input: $\left(n, P, m=\frac{n}{a}-1,t^*,\ket{\psi(0)},X\right)$, the number of nodes $n$ of the graph $\mathcal{G}$, the transition matrix $P$ of the graph $\mathcal{G}$, the linear relationship $m^*=\left[\frac{n}{a}-1\right]$ (Eq.(\ref{eq_linear_relationship_numerical})), the running time $t^*$, the initial quantum state for the quantum walk $\ket{\psi(0)}$ and the position operator. Output: A truth value, $\mathcal{C}$, concerning the completeness of the graph $\mathcal{G}$.}
		\label{fig:Flowchart}
\end{figure}	
The goal of our algorithm is to determine whether the graph $\mathcal{G}$ is complete and it can be articulated in two stages, the algorithm workflow is as follows:
\begin{itemize}
	\item{\textbf{Stage 1: Ruling-out}}\\Focuses on reliably excluding highly sparse graphs and has a high negative predictive value. A negative result in this phase confidently indicates the absence of completeness in the graph examined. This phase uses a quantum walk based measurement and its computational complexity is low.
	\item{\textbf{Stage 2: Verification}}\\Handles the positive cases from Stage 1. This phase is meant to use a more computationally intensive method, quantum phase estimation (QPE). It serves to confirm the positive result of the first part, which is not reliable if not verified.
\end{itemize}
We use the Szegedy quantum walk with marked vertices, leveraging the linear relationship between $n$ and $m$ (Eq.~(\ref{eq_linear_relationship_numerical})) explored in Section \ref{sub_II_C}. Through this relationship, we calculate $m^*$, representing the optimal number of marked nodes $M=\{x_1,\dots,x_{m^*}\}$ for a complete graph. We mark randomly $m^*$ nodes on the graph $\mathcal{G}$ to ensure the coincidence of the hitting time $t_{h}$ and the maximum $t_{max}$ of the probability $P_{M}(t,n,m)$, resulting in the modified transition matrix $P'$.

This marking maximizes the probability of being on a marked node in the shortest time possible, assuming $\mathcal{G}_c$ is a complete graph. We evolve the initial state $\ket{\psi(0)}$ (Eq.~(\ref{eq_initial_state_complete_case})) with the operator $W_{P'}$ for a running time $t^*$, chosen to be the closest integer to the second maximum of the probability $P_M(t,n,m)$, which is $t^*=3$. The choices of the running time $t^*$ and the number of marked nodes $m^*$ ensure that the probability $P^*_M(t^*)$ is close to $1$ for a complete graph (see Sec.~\ref{sub_II_E}). The structure of the algorithm is as follows:

\begin{center}
    \textbf{Stage 1: Ruling-out}
\end{center}

\textbf{Input:}
The transition matrix $P$ of an undirected graph $\mathcal{G}$, the total number of nodes $n$ of $\mathcal{G}$, the linear relationship in Eq.~(\ref{eq_linear_relationship_numerical}), the running time $t^*=3$, the initial state $\ket{\psi(0)}$ (Eq.~(\ref{eq_initial_state_complete_case})) and the position operator $X$.

\textbf{Output:}
A truth value $\mathcal{C}$ indicating the completeness of the graph.

\begin{enumerate}
    \item[\textbf{1.}] Evaluate $m^*$ corresponding to $n$ using the linear relationship (\ref{eq_linear_relationship_numerical}) and rounding to the closest integer: $m^* =\left[(n-1)/a\right]$.
    
    \item[\textbf{2.}] In the graph $\mathcal{G}$, mark $m^*$ randomly selected nodes. Let $M \subset V$ be the subset of marked nodes, where $V = \{1,\dots, n\}$ and $M = \{x_1, \dots, x_{m^*}\}$. Update the transition matrix to $P'$.
    
    \item[\textbf{3.}] Given the transition matrix $P'$, run the Szegedy quantum walk for a time $t^*$ by applying the evolution operator $W_{P'}^{t^*}$ to the initial state $\ket{\psi(0)}$.
    
    \item[\textbf{4.}] At the end of the evolution, measure the position of the walker:
    \begin{gather*}
        X\ket{\psi(t^*)} = \ket{x} \qquad \text{where} \qquad 
        \begin{cases}
            x \in M \\
            x \notin M
        \end{cases}
    \end{gather*}
	\begin{itemize}
        \item If $x \notin M$, then the graph is not complete, $\mathcal{C} = \text{False}$, and the algorithm terminates.
        \item If $x \in M$, proceed to Stage 2.
    \end{itemize}
\end{enumerate}
Several comments are in order for this first part of the algorithm.
Given that the probability $P^*_{M}(t^*)$ of finding the walker at a marked node at time $t^*$ is close to $1$ for a complete graph with $n$ nodes and $m^*$ marked ones, if we find the walker at an unmarked node, we can conclude that the graph is not complete. As noted in Sec.~(\ref{sub_II_E}), provided that $P^*_{M}(t^*) \approx 1$ for a complete graph:
\begin{itemize}
    \item Not finding a marked node $\Rightarrow$ the graph is incomplete.
\end{itemize}
However, it is also true that:
\begin{itemize}
    \item Finding a marked node $\nRightarrow$ the graph is complete.
\end{itemize}
This is because, even if the graph is \textit{not complete}, the probability of finding the walker at a marked node is non-zero.

Therefore, if we find an unmarked node, we conclude the algorithm and determine that the graph $\mathcal{G}$ is incomplete. If we do find a marked node, we proceed to the QPE subroutine. We use QPE \cite{kitaev1995quantum, NielsenAndChuang} in the second stage to estimate one of the eigenvalues' phases of the operator $W_{P'}$, which we denote as $\theta_j$ without loss of generality. We compare the estimated $\theta_j$ with $\theta_2$, the eigenphase that we expect to find in the complete case. Before proceeding, since for the QPE subroutine we do not need the number of marked nodes $m^*$ defined through the linear relationship, we change $m = 1$, and, in the graph $\mathcal{G}$, we mark only one arbitrary node, updating the transition matrix to $P''$ and consequently preparing the correspondent evolution operator $W_{P''}$. We calculate analytically the value of the phase $\theta_2$ using $\cos(\theta_2) = \frac{n-m-1}{n-1}$. Choosing this number of marked nodes minimizes the loss of information on the transition matrix $P$ in the process of marking nodes. Let $\ket{\theta_2^+}$ be the eigenvector of $W_{P''_c}$ associated with the phase $\theta_2$, which is defined in terms of the computational basis of $H^{n^2}$ in \cite{Portugal_book} and can be prepared based on \cite{Zalka, Efficient_QC_for_SQW_loke}. The structure of the second stage is as follows:

\begin{center}
    \textbf{Stage 2: Verification}
\end{center}

\textbf{Input:} 
The initial state $\ket{0}^{\otimes p}\ket{\theta_2^+}$ and the operator $W_{P''}$. The integer $p$ represents the number of qubits in the first of the two qubit registers used by the algorithm, corresponding to the precision $2^p$ digits of the phases that the algorithm will estimate. The state $\ket{\theta_2^+}$ in the second register is the eigenstate corresponding to the eigenphase $\theta_2$ of the evolution operator $W_{P''_c}$ of the Szegedy quantum walk on a complete graph with $n$ nodes and $m^*$ marked ones.

\textbf{Output:} 
Estimate of one of the eigenphases $\theta_j$ of $W_{P''}$.

\begin{enumerate}
	\item[\textbf{5.1}] Change the number of marked nodes $m$ to $1$ and update the transition matrix from $P$ to $P''$.
	\item[\textbf{5.2}] Prepare the operator $W_{P''}$.
    \item[\textbf{5.3}] Prepare the initial state $\ket{0}^{\otimes p}$ in the first register and $\ket{\theta_2^+}$ in the second register.
    \item[\textbf{5.4}] Apply a Hadamard gate $H$ to each qubit of the first register.
    \item[\textbf{5.5}] Apply $W_{P''}^{2^{0}}, W_{P''}^{2^1}, \dots, W_{P''}^{2^{p-1}}$ to the second register, each controlled by one of the $p$ qubits of the first register.
    \item[\textbf{5.6}] Apply the inverse quantum Fourier transform \cite{coppersmith2002approximate, NielsenAndChuang} to the first register.
    \item[\textbf{5.7}] Measure the first register in the computational basis.
    \item[] \textbf{Output:} \\
    Estimate of $\theta_j$.
\end{enumerate}

The final steps of the algorithm, after the QPE subroutine, are:

\begin{enumerate}
    \item[\textbf{6.}] Compute the phase $\theta_2$ from $\cos(\theta_2) = \frac{n-m-1}{n-1}$, with $m=1$. 
	\item[\textbf{7.}] Take the returned value $\theta_j$ and compare it with $\theta_2$:
    \begin{itemize}
        \item If $\theta_j = \theta_2$, then the graph $\mathcal{G}$ is complete, $\mathcal{C} = \text{True}$.
        \item If $\theta_j \neq \theta_2$, then the graph $\mathcal{G}$ is not complete, $\mathcal{C} = \text{False}$.
    \end{itemize}
\end{enumerate}

Further comments are in order on the second stage of the algorithm. We will obtain the eigenphase $\theta_j = \theta_2$ if and only if the graph is complete. The choice of the number of qubits in the first register of the QPE subroutine is related to the desired precision for the estimation of the eigenphase $\theta_j$, which is related to the size of the graph $\mathcal{G}$ and affects the complexity of the second stage and of the entire algorithm. Thus, the total complexity is the sum of the QPE subroutine complexity for the second stage, and $t^*$ for the first stage. A more detailed analysis of this complexity is presented in Subsection \ref{sub_III_B}.

\subsection{Complexity Analysis}\label{sub_III_B}

In this section, we analyze the complexity of the two stages of the algorithm, discussing the contribution of the QPE subroutine using the big-O notation.

As discussed in Section \ref{sub_II_E}, Stage $1$ of the algorithm has a running time of $t^*$, corresponding to the second maximum of the probability function $P_M(t,n,m)$. According to Eq.(\ref{eq_tmax_second}), this running time is independent of $n$ when we use the linear relationship in Eq.(\ref{eq_linear_relationship_numerical}). Therefore, the running time of the first part of the algorithm is $t^* = 3$ quantum walk steps, making the time complexity of the entire algorithm, in the worst case, dependent solely on Stage $2$ and thus on the phase estimation.

In the Verification stage we employ QPE aided with classically computed variables. Well-known algorithms using the QPE subroutine, such as those for period finding and prime factorization \cite{NielsenAndChuang, Quantum_counting_AA, kitaev1995quantum}, achieve efficiency in applying controlled$-U^j$ gates and thus the complexity of the whole QPE application is based entirely on the inverse Quantum Fourier Transform (iQFT) in the QPE. This is because of the inherent characteristics of these specific problems, rather than due to the optimization of the controlled$-U^j$ gates themselves. In our work we consider the application of $C-W_{P''}^j$ gates efficient, compared to the iQFT, in order to dedicate our attention to an intriguing effect on the this latter component. In particular, the QFT complexity transforms from its usual quadratic complexity to polylogarithmic behavior. To define the complexity of the iQFT we must determine the number $p$ of qubits in its first register, which corresponds to the number of bits of precision in the QPE eigenphase measurement. This number $p$ sets the running time for the inverse quantum Fourier transform, which is $\mathcal{O}(p^2)$.

Estimating $p$ involves determining the number of bits required to decide if the phase $\theta_j$ measured with the QPE is equal to $\theta_2$, when the graph structure is as close as possible to the complete graph, representing the worst case scenario. The detailed study for deriving this number $p$ is presented in Appendix \ref{Appendix_B}, supporting the following proposition:
\begin{prop}{QPE subroutine complexity.\\}
	The number of qubits $p$ can be calculated based on a lower bound function $\mathcal{F}(n)$ for the difference between the phases $\theta_j$ and $\theta_2$
	leading to:
	\begin{gather}
		p \approx \left|\log_2\left(\mathcal{F}(n)\right)\right|+1 \sim \log_2(n)+1.
	\end{gather}
	 Thus, the complexity of the QPE subroutine is:
	 \begin{gather}
	 	\mathcal{O}(\log^2n).
	 \end{gather}
	\qed
\end{prop}

Indeed, provided that the controlled-$W_{P'}$ can be efficiently implemented, the time complexity of the QPE subroutine is determined by the inverse quantum Fourier transform, which involves $\mathcal{O}(p^2)$ steps. Consequently, the complexity is $\mathcal{O}(\log^2 n)$, polylogarithmic in $n$. Since we are analyzing the worst case scenario, we suppose obtaining from the first part of the algorithm the result $x \in M$ (see step $4$ of the algorithm in Sec.\ref{sub_III_A}), and therefore proceeding to Stage $2$. As a result, the total complexity of the algorithm in the worst case coincides with that of the QPE subroutine since we only need to add the constant complexity of the first part.

To compare the complexity of our algorithm with a classical one, we consider the number of steps required to check all the links in a graph $\mathcal{G}$, which is $\mathcal{O}(n^2)$. 
We want to point out that our algorithm is inherently different from a pure search algorithm, and thus differs from a Grover algorithm. For example, a Grover search might include an oracle that can highlight missing links in the incomplete graph and then identify them by a Grover search, leading to a complexity $O(n)$. In our algorithm, we do not use such an oracle and we are not searching for marked elements either. Our algorithm uses properties of quantum walk search in order to deduce information on the graph structure, while combining it with QPE. Finally, since our algorithm provides the eigenvalue $\theta_j$ of the operator $W_{P'}$ we can compare it with the classical eigenvalue QR algorithm \cite{QR_algorithm,QR_algorithm_2} which complexity $\mathcal{O}(n^3)$ we evidently outperform in our hypotheses. The complexity comparison is summarized in Table \ref{table_complexity}.

\begin{table}[t]
	\centering
	\begin{tabular}{ll}
		\hline
		\multicolumn{1}{l}{\textbf{Algorithm}}                                                           & \multicolumn{1}{c}{\textbf{Complexity}}             \\ \hline
		\multicolumn{1}{l}{Classical Completeness Testing Algorithm}      & \multicolumn{1}{c}{$\mathcal{O}(n^2)$} \\ \hline
		\multicolumn{1}{l}{QR Eigenvalue Algorithm}                       & \multicolumn{1}{c}{$\mathcal{O}(n^3)$} \\ \hline
		\multicolumn{1}{l}{Quantum Completeness Testing Algorithm}        & \multicolumn{1}{c}{$\mathcal{O}(\log^2n)$} \\ \hline
	\end{tabular}
	\caption{Comparing the complexity of the quantum completeness testing algorithm with classical and semi-quantum approaches.}
	\label{table_complexity}
\end{table}

\section{Applications of Graph Completeness Testing}\label{sec_IV}

In the following we provide a number of examples where the quantum algorithm for testing the completeness of a graph can solve problems of practical utility:
\paragraph{Network Connectivity in Computer Science\\}
\textit{Description}: Graph completeness is crucial in network connectivity, ensuring all nodes are interconnected. This is important for designing efficient network routing algorithms and understanding the overall structure of networks. 
\textit{Key Use}: Enhancing the design and performance of network routing algorithms by ensuring full connectivity.\\
\textit{Example: Dynamic Routing Optimization.}
In dynamic routing \cite{Dynamic_routing}, the network topology significantly influences routing decisions. Therefore, rapid and efficient completeness checking enhances the performance of any algorithm in this category, facilitating efficient data transmission.

\paragraph{Graph Traversal Algorithms\\}
\textit{Description}: These algorithms are used to explore or visit all nodes in a graph systematically \cite{Graph_algorithms,Introduction_to_algorithms}. Knowing whether a graph or its subparts are complete helps optimize traversal processes.
\textit{Key Use}: Applications in pathfinding, network analysis, and data mining where completeness ensures efficient exploration and avoids unnecessary computations. 
\textit{Benefit}: Pre-evaluating graph completeness aids in estimating the time complexity of traversal algorithms and enhancing their performance.\\
\textit{Example: Graph Traversal Optimization.}
When executing graph traversal algorithms such as Depth-First Search (DFS) or Breadth-First Search (BFS) \cite{Graph_algorithms,Introduction_to_algorithms}, recognizing that a graph is complete can prevent redundant connectivity checks, thereby accelerating the process.

\paragraph{Network Clustering\\}
\textit{Description}: This involves grouping nodes in a network based on connectivity patterns. Completeness testing identifies fully connected clusters, aiding in recognizing communities or subgroups. 
\textit{Key Use}: Social network analysis, community detection, and network visualization. 
\textit{Benefit}: Helps in identifying tightly-knit clusters which are essential for accurate analysis and visualization.\\
\textit{Example: Community Detection in Social Networks.}
Completeness testing assists in identifying subgraphs where every member is connected to every other member, thereby facilitating the detection of tight-knit communities \cite{Community_detection_twitter,Community_detection_comparative}.

\paragraph{Network Partitioning\\}
\textit{Description}: Dividing a network into smaller, fully connected subnetworks or partitions or examining partitions of a pre-existent network. Completeness testing ensures that partitions are fully connected. 
\textit{Key Use}: Applications in load balancing, fault tolerance, network management, and in improving the performances of clique finding algorithms.
\textit{Benefit}: Optimizes the performance of partitioning algorithms by ensuring efficient and reliable subnetworks.\\
\textit{Example: Clique Finding.}
Completeness testing can significantly reduce the complexity of clique finding algorithms \cite{Clique_problem}, eliminating the need for sequential checks over the network's edges.

\paragraph{Fairness Checking in Pairwise Comparison Algorithms\\}
\textit{Description}: Ensures all nodes are compared with each other in algorithms requiring pairwise comparisons. Completeness is crucial for fairness in such processes. 
\textit{Key Use}: Voting systems and decision-making processes, which determines winners based on pairwise comparisons. 
\textit{Benefit}: Ensures no node (or candidate) is left out of comparisons, maintaining fairness and accuracy in results.\\
\textit{Example: Voting Systems (Copeland Method).\\}
In the Copeland voting system, completeness ensures that every candidate is compared with all others, guaranteeing a fair election process \cite{Copeland1,Condorcet1}.

\paragraph*{}
Incorporating graph completeness testing can enhance the efficiency, accuracy, and fairness of these applications in their respective domains. Furthermore, beyond improving the performance of existing classical algorithms, graph completeness testing can be utilized in purely quantum algorithms, including quantum network routing and quantum community detection \cite{Cade_community_detection_quantum}, among others.

\section{Conclusions}\label{sec_V}

In this work, we have presented a novel quantum algorithm for testing the completeness of a graph, leveraging the Szegedy quantum walk and the QPE subroutine. Our findings demonstrate the algorithm's efficiency in determining whether a graph is complete.
Key contributions and results of our work include:

    \noindent \textit{1.~Linear Relationship Between Marked and Total Nodes}:~We discovered a linear relationship between the number of marked nodes \( m \) and the total number of nodes \( n \) in a complete graph. This relationship ensures that the hitting time \( t_h \) coincides with the maximum of the probability function \( P_M(t,n,m) \). This allows us to determine the optimal number of marked nodes \( m^* \), guaranteeing a high probability of locating the walker on a marked node in a complete graph.

    \noindent \textit{2.~Constant Maximum Probability Time}:~By introducing this linear relationship, we showed that the time \( t_{max} \), at which the probability of finding the walker on a marked node is maximized, remains constant and independent of the number of nodes in the graph. This insight enables the first stage of our completeness checking algorithm to operate with a constant time complexity.

    \noindent \textit{3.~Quantum Phase Estimation Subroutine Complexity}:~To address the insufficiency of the initial condition for determining graph completeness, we incorporated the QPE in the second stage of the algorithm in an original fashion. We demonstrated that the number of qubits of the first register in the QPE is given by $ p \sim \log_2\left( n\right) + 1 $, derived from the lower bound of the eigenphase difference between complete and incomplete graphs. QPE exhibits a complexity of \( \mathcal{O}(\log^2 n) \), given the hypotheses in \ref{sub_III_B}, offering a more efficient alternative to classical algorithms for graph completeness testing.

    \noindent \textit{4.~Comparison with Grover Algorithm}:~We showed that our algorithm outperforms Grover-based algorithms for graph completeness testing, both in terms of efficiency and complexity.

    \noindent \textit{5.~Practical Applications}:~The proposed completeness testing algorithm has significant implications for various applications, including network connectivity, graph traversal, network clustering, network partitioning, and fairness checking.

In this work, our first major result examines the significance of the optimal number of marked nodes, as derived from a linear relationship, Eq.(\ref{eq_linear_relationship_numerical}). Here, we consider the inverse viewpoint: instead of fixing the total number of nodes and deriving the number of marked nodes from the linear relation (as done throughout the work), we fix the marked cluster dimension and from the linear relationship we can deduce the total dimension of the graph.

To explain this from an intuitive perspective, let us envision a graph-like environment where nodes can be added or removed as needed, subject to the constraint of some special nodes grouped together. Consider a classical walker navigating the nodes by following the edges of a complete graph. The objective is to configure the environment’s layout, adjusting the number of nodes, to maximize the probability that the walker lands on the special cluster. In this classical scenario, the solution is straightforward: eliminate all nodes that are not part of the special cluster. However, the scenario changes when dealing with a quantum walker, as explained through Szegedy's quantum walk evolution. In the quantum case, the problem becomes more complex. Unlike classical random walks that converge to a stationary distribution as time approaches infinity, quantum walks result in a probability distribution over the nodes that varies periodically with time. Therefore, identifying the optimal time to maximize the probability of landing on the special cluster is crucial. To maximize the probability of the walker being on the special cluster of nodes within the shortest amount of time, as outlined in \textbf{Proposition \ref{solution_of_the_optimality_condition}} in Sec.\ref{sub_II_D}, the entire graph must be slightly larger than just the cluster. This requirement arises solely from the quantum nature of the Szegedy Quantum Walk that governs the walker’s evolution on the complete graph.

Our quantum algorithm for graph completeness testing provides a robust and efficient method, marked by its novel use of the Szegedy quantum walk and the QPE. This work not only advances the field of quantum computing but also offers practical solutions for real-world graph analysis problems.

	\section*{Acknowledgements}
		The authors acknowledge support from Spanish MICIN grant PID2021-122547NB-I00 and the “MADQuantum-CM"project funded by Comunidad de Madrid and by the Recovery, Transformation and Resilience Plan – Funded by the European Union - NextGenerationEU, the CAM/FEDER Project No. S2018/TCS-4342 (QUITEMAD-CM), M.A. M.-D. has been partially supported by the U.S.Army Research Office Through Grant No. W911NF-14-1-0103. S.G. acknowledges support from a QUITEMAD grant. This work has been financially supported by the Ministry for Digital Transformation and of Civil Service of the Spanish Government through the QUANTUM ENIA project call – Quantum Spain project, and by the European Union through the Recovery, Transformation and Resilience Plan – NextGenerationEU within the framework of the Digital Spain 2026 Agenda.
	
	\appendix
\section{Analytical Solution of the Optimality Condition}
\label{Appendix_A}

In this appendix, we detail the analytical derivation of the Solution of the Optimality Condition \eqref{solution_of_the_optimality_condition_eq} shown in Sec.~\ref{sub_II_C}. To solve the system \eqref{system_F(T)_tmax} and obtain the linear relationship, we start by writing the explicit trigonometric form of the Chebyshev polynomial of the first kind present in Eq.(\ref{eq_F(T)_complete_case}):
\begin{gather}
    U_{2T}\left[\frac{n-m-1}{n-1}\right] = \frac{\sin \left((2T+1) \arccos\left(\frac{-m+n-1}{n-1}\right)\right)}{\sqrt{1 - \left(\frac{n-m-1}{n-1}\right)^2}}.
\end{gather}
Substituting the expression for \(T\) from Eq.(\ref{eq_t_max_analytical}), we simplify to obtain:
	\begin{gather}
		\Scale[0.9]{U_{2T}{\left[\frac{n-m-1}{n-1}\right]}=\frac{\sin \left(\arccos\left(1-\frac{m}{n-1}\right)+\arctan\left(\sqrt{-\frac{m-2 n+2}{m}}\right)\right)}{\sqrt{2-\frac{m}{n-1}} \sqrt{\frac{m}{n-1}}}}.
	\end{gather}
Using the addition formula for the sine function,  we can simplify it further:
\begin{gather}
    U_{2T}\left[\frac{n-m-1}{n-1}\right] = \sqrt{\frac{n-1}{2m}}.
\end{gather}
Substituting this expression into the first equation of the system \eqref{system_F(T)_tmax}, we derive the following simplified equation:
\begin{gather}
    \frac{m}{n} + \frac{2 (n-1)(n-m) \left(-\sqrt{\frac{n-1}{2m}} + \gamma + 1\right)}{n (2n-m-2) \left(\frac{1}{2}\gamma + 1\right)} - 1 = 0,
\end{gather}
where \(\gamma\) is defined as:
\begin{gather}
    \gamma := \frac{\arctan\left(\frac{\sqrt{2n-m-2}}{\sqrt{m}}\right)}{2\arctan\left(\frac{\sqrt{m}}{\sqrt{2n-m-2}}\right)}.
\end{gather}
To simplify \(\gamma\), we use the trigonometric equivalence:
\begin{gather}
    \arccos(x) = 2 \arctan\left(\frac{\sqrt{1-x^2}}{x+1}\right),
\end{gather}
which holds for \(-1 < x < 1\). Setting \(x = \frac{n-m-1}{n-1}\), we find \(x < 1\) if \(n > 1\), \(m > 0\), and \(n > m\), suitable assumptions for a graph with \(n\) nodes and \(m\) marked nodes. Requiring \(n > \frac{m}{2} + 1\) we ensure $x>-1$, which is consistent with our educated ansatz to be introduced later on. Substituting the equivalence into the denominator of \(\gamma\), we get:
\begin{gather}
    \arccos\left(1 - \frac{m}{n-1}\right) = 2 \arctan\left(\frac{\sqrt{m}}{\sqrt{2n-m-2}}\right).
\end{gather}
This substitution gives:
\begin{gather}
    \gamma = \frac{\arctan\left(\frac{\sqrt{2n-m-2}}{\sqrt{m}}\right)}{2 \left(\frac{\pi}{2} - \arctan\left(\frac{\sqrt{2n-m-2}}{\sqrt{m}}\right)\right)}.
\end{gather}
Substituting \(\gamma\) into the main equation, we obtain:
\begin{gather}\label{eqA_main_eq}
    \frac{(m - n) \quad \mathcal{Q}(n, m)}{n (2n-m-2) \left(2 \pi - 3 \arctan\left(\sqrt{\frac{2n-m-2}{m}}\right)\right)} = 0,
\end{gather}
where \(\mathcal{Q}(n, m)\) is:
	\begin{gather} \notag
		\Scale[0.95]{\mathcal{Q}\left(n,m\right)=2 \pi  \left(m-\sqrt{2} m \left(\frac{n-1}{m}\right)^{3/2}\right)+}\\
		\Scale[0.95]{+\left(2 (n-1) \left(2 \sqrt{2} \sqrt{\frac{n-1}{m}}+1\right)-3 m\right) \arctan\left(\sqrt{-\frac{m-2 n+2}{m}}\right).\label{eq_classical_quantum_solution}}
	\end{gather}
Equation (\ref{eq_classical_quantum_solution}) is satisfied by \(m = n\) and by \(\mathcal{Q}(n, m) = 0\). The first solution, \(m = n\), represents a trivial solution to the optimality problem of maximizing \(P_M(t, n, m)\) when \(t = t_{max}\). To find the second solution, we solve \(\mathcal{Q}(n, m) = 0\) for \(m\).
	\begin{gather}\notag
		\Scale[0.9]{2 \pi  \left(m-\sqrt{2} m \left(\frac{n-1}{m}\right)^{3/2}\right)+}\\
		\Scale[0.9]{+\left(2 (n-1) \left(2 \sqrt{2} \sqrt{\frac{n-1}{m}}+1\right)-3 m\right) \arctan\left(\sqrt{-\frac{m-2 n+2}{m}}\right)=0.}
	\end{gather}	
This transcendental equation in \(m\) cannot be solved analytically. We use a suitable ansatz for the solution, suggesting a linear relationship between \(n\) and \(m\). Substituting \(n = am + 1\) into the equation, we obtain:
	\begin{gather}\label{eqA_a_equation}
		\Scale[0.9]{\left(4 \sqrt{2} a^{3/2}+2 a-3\right) \arctan\left(\sqrt{2 a-1}\right)-2 \pi  \left(\sqrt{2} a^{3/2}-1\right)=0}
	\end{gather}
This equation, with \(a\) as the only variable, confirms the linear relationship between \(n\) and \(m\). Solving numerically, we find \(a \approx 1.44512\). Substituting this value, we obtain the linear relationship solution 
to the Optimality Condition (OC):
\begin{gather}
    n = 1.44512 m + 1.
\end{gather}
This relationship allows us to calculate the \textit{optimal} number of nodes \(m^*\) to mark in a graph with \(n\) nodes, ensuring \(t_h\) coincides with the first maximum of \(P_M(t, n, m)\), thereby maximizing the probability of being on a marked node in the shortest possible time.

To ensure real solutions from Eq.~\eqref{eqA_main_eq}, we require the inverse trigonometric functions to yield real values, which hold if \(n > m\) and \(n > 1\), consistent assumptions for a graph with \(n\) nodes and \(m\) marked nodes. Additionally, ensuring the denominator of Eq.(\ref{eqA_main_eq}) is non-zero, we substitute \(n = am + 1\):
\begin{gather}
    (2a - 1)(am + 1)\left(2\pi - 3 \arctan\left(\sqrt{2a - 1}\right)\right) \neq 0.
\end{gather}
Since \(am + 1 = n \neq 0\), the inequality holds for \(a \neq \frac{1}{2}\) and \(a \neq \frac{1}{2} + \frac{1}{2} \tan^2\left(\frac{2\pi}{3}\right)\), or \(a \neq 2\).

\section{Number of bits for the QPE }
\label{Appendix_B}
We observe from Eq.(\ref{eq_eigenphases_complete_case}) and Eq.(\ref{eq_chain_inequal_spectral_radius}) that for a complete graph with $n$ nodes and $m$ marked nodes, with adjacency matrix $A_{_c n-m}$, we have:
\begin{gather}
    \cos(\theta_2) = \frac{n-m-1}{n-1} = \frac{\rho(A_{_c n-m})}{n-1} = \frac{\sqrt{2|E_c| - n + 1}}{n-1}.
\end{gather}
For a not complete graph with the same number of nodes and marked nodes, and adjacency matrix $A_{n-m}$, we have:
\begin{gather}
    \lambda < \rho(A_{n-m}) < \sqrt{2|E| - n - 1},
\end{gather}
where $\lambda$ is an eigenvalue of $A_{n-m}$ \cite{Bound_spectral_radius}. With a suitable normalization factor $\mathcal{N}$, we find:
\begin{gather}
    \cos(\theta_j) = \frac{\lambda}{\mathcal{N}} < \frac{\rho(A_{n-m})}{\mathcal{N}} < \frac{\sqrt{2|E| - n + 1}}{\mathcal{N}},
\end{gather}
where $\theta_j$ is the eigenphase corresponding to the eigenvalue $\lambda$. Although we do not have an explicit expression for $\mathcal{N}$ in the incomplete case, we assume a normalizing ansatz $\tilde{\mathcal{N}}$ such that $\tilde{\mathcal{N}} < \mathcal{N}$. Hence:
\begin{gather}
    \cos(\theta_j) < \frac{\sqrt{2|E| - n + 1}}{\mathcal{N}} < \frac{\sqrt{2|E| - n + 1}}{\tilde{\mathcal{N}}}.
\end{gather}
Applying the inverse cosine function and subtracting the positive phase $\theta_2$ from both sides of the inequality we obtain:
\begin{gather}\label{eq_thetaj_-_theta2}
    \theta_j - \theta_2 > \arccos{\left(\frac{\sqrt{2|E| - n + 1}}{\tilde{\mathcal{N}}}\right)} - \theta_2.
\end{gather}
This inequality provides a lower bound for the difference between the phase $\theta_j$ (for a not complete graph) and the phase $\theta_2$ (for a complete graph). This lower bound, once converted to binary representation, helps us determine how many bits of the binary conversion of $\theta_j$ we need to estimate with the QPE to distinguish it from binary $\theta_2$. Let us call binary $\theta_j$ and $\theta_2$ respectively $\theta_j^b$ and $\theta_2^b$. If $\theta_j^b$ and $\theta_2^b$ are equal up to the $l$-th bit, then: $(\theta_j - \theta_2)^b = \theta_j^b - \theta_2^b$ will have all zeros up to the $(l+1)$-th bit. Renaming
\begin{gather}\notag
	\left(\arccos{\left(\frac{\sqrt{2|E| - n + 1}}{\tilde{\mathcal{N}}}\right)}\right)^b = \tilde{\theta}_j^b,
\end{gather} Eq.(\ref{eq_thetaj_-_theta2}) becomes:
\begin{gather}\label{eq_thetaj_-_theta2_binary}
    (\theta_j - \theta_2)^b > (\tilde{\theta}_j - \theta_2)^b.
\end{gather}
If $\tilde{\theta}_j^b$ and $\theta_2^b$ are equal up to the $f$-th bit, their difference will have all zeros up to the $(f+1)$-th bit. By construction $f$ will always be greater than $l$. Therefore, $f$ serves as an upper bound for $l$, and we need to estimate at most $f+1$ bits of $\tilde{\theta_j}^b$ with the QPE to distinguish it from $\theta_2^b$, hence this holds also for $\theta_j^b$.

Since choosing a suitable $\tilde{\mathcal{N}}$ is not straightforward, we fit some simulated $\theta_j - \theta_2$ data (see Appendix \ref{Appendix_C} for more details) with respect to $n$ and adjust the fitting function $\mathcal{F}(n)= \tilde{\theta}_j - \theta_2$ to serve as a lower bound for those data. Eq.(\ref{eq_thetaj_-_theta2_binary}) becomes:
\begin{gather}\label{eq_thetaj_-_theta2_binary_2}
	 (\theta_j - \theta_2)^b >\mathcal{F}(n)^b,
\end{gather}
where $\mathcal{F}(n)^b$ is the binary conversion of $\mathcal{F}(n)$. The function $\mathcal{F}(n)$ will serve as a lower bound independently of the graph's sparsity, since we are examining the worst case scenario of an almost complete graph. Therefore, we utilize $\mathcal{F}(n)$ to calculate the number of bits $f$ needed. The binary version of $\mathcal{F}(n)$ is a non integer binary number with zero whole part. Assuming to evaluate it up to $g$ bits in total, we can express it as $0.$ followed by $f$ zeroes, then the rest of the fractional part up to $g$ bits: $\mathcal{F}(n)^b=0.0_1 0_2 \dots 0_f \ x_{f+1} \ x_{f+2} \dots x_{g}$, where $x_i \in \left\{0,1 \right\} $. This means that $\mathcal{F}(n)$ is larger than or equal to $2^{-f}$, and smaller than $2^{-f+1}$, hence the logarithm in base 2 of $\mathcal{F}(n)$ is larger or equal to $-f$ and smaller than $-f+1$. Which, in turn, means that the whole part of the $\log_2$ of $\mathcal{F}(n)$ is approximately equal to $-f$. Therefore, we can calculate the number of bits $f$:
\begin{gather}\label{eq_bitwise_comparison_log2}
    f \approx |\log_2(\mathcal{F}(n))|.
\end{gather}

	\begin{figure}[t]
		\centering
		\includegraphics[width=1\columnwidth]{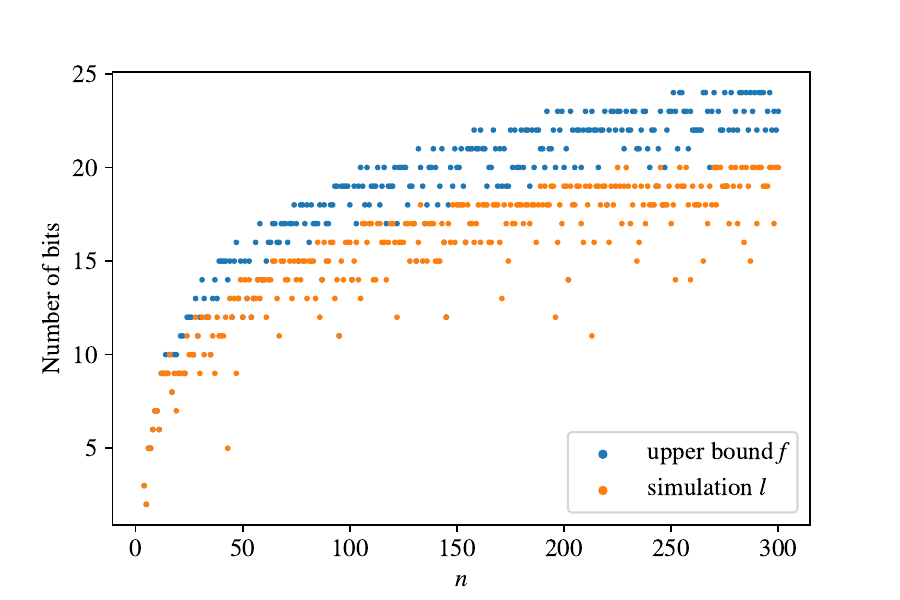}
		\caption{Comparison between the number of bits $f$ and $l$ that we need to evaluate in the QPE. The number of bits $f$ is calculated from $\mathcal{F}(n)$, while the number of bits $l$ is calculated from the simulated data.}
		\label{fig_f_vs_l}
	\end{figure}
	\begin{figure}[t]
		\centering
		\includegraphics[width=1\columnwidth]{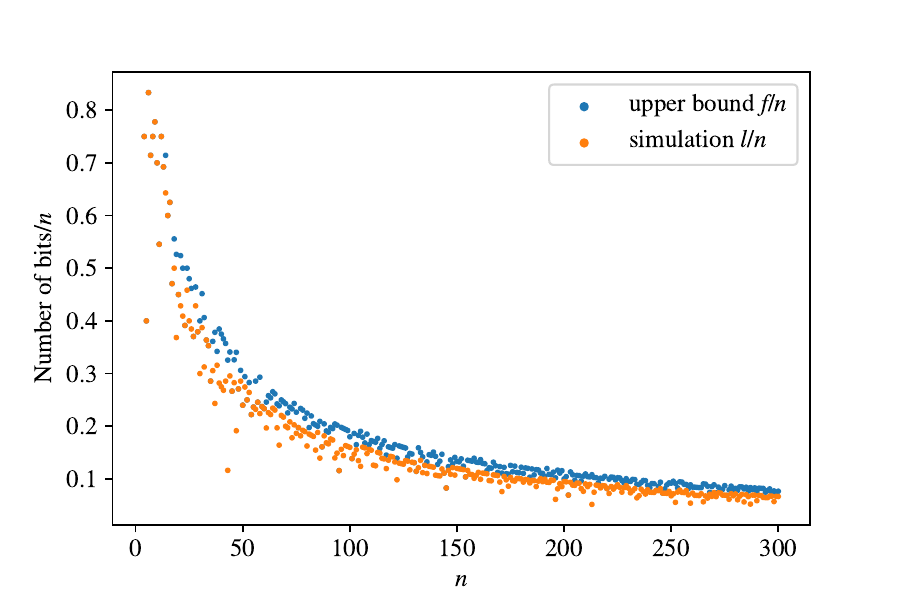}
		\caption{Comparison between the ratio $f/n$ and $l/n$. The number of bits $f$ is calculated through \ref{eq_F(n)}, while the number of bits $l$ is calculated from the simulated data. There is a clear trend for the ratio $f/n$ and for the ratio $l/n$ to be always less than $1$.}
		\label{fig_f_vs_l_over_n}
	\end{figure}
Consequently, the number of qubits $p$ needed in the first register of the QPE subroutine is:
\begin{gather}
    p = f + 1 \approx |\log_2(\mathcal{F}(n))| + 1.
\end{gather}
The fitting function for $\mathcal{F}(n)$ is obtained from simulated data in Appendix \ref{Appendix_C} yielding the following scaling behaviour:
\begin{gather}\label{eq_F(n)}
    \mathcal{F}(n) \sim \frac{13}{n^{3.4}}.
\end{gather}
Therefore:
\begin{gather}
    p \sim \left| \log_2\left(\frac{13}{n^{3.4}}\right) \right| + 1.
\end{gather}
Further simplifying this expression defines $p$ as:
\begin{gather}
	p \sim \left|3.4\log_2\left(13 n\right)\right| +1.
\end{gather}
The number of qubits $p$ enters in the QPE complexity, determining the time complexity of the inverse quantum Fourier transform $\mathcal{O}\left(p^2\right)$, resulting in $\mathcal{O}\left(\log^2n\right)$ for the complexity of the QPE subroutine. Fig.\ref{fig_f_vs_l} shows how the upper bound $f$ always exceeds the simulation data $l$, while Fig.\ref{fig_f_vs_l_over_n} shows that the ratios $f/n$ and $l/n$ are always less than $1$, indicating that the number of bits needed in the QPE is always less than the number of nodes in the graph. This property conveys that the size of the quantum circuit needed to perform the QPE never reaches the size of the system we are analyzing, moreover, it does not scale as this latter either.

\section{Szegedy quantum walk simulation}
\label{Appendix_C}
 
In this appendix we provide a detailed description of the Szegedy quantum walk simulation used to generate the data needed to estimate the number of bits $l$ and $f$ for the QPE subroutine. We simulate data related to the worst case scenario, where the graph is incomplete while its dynamical eigenphase $\theta_j$ remains close to the value $\theta_2$ of the complete case, thus estimating the highest number of bits needed in the QPE to distinguish the two phases. To this end, we compute the quantum walk evolution operator associated with $\mathcal{G}$ with $n$ nodes, $m=1$ marked node and only one link less than a complete graph $\mathcal{G}_c$ with the same number of marked and unmarked nodes. We calculate the eigenvalues and eigenvectors of the evolution operator $W_{P'}$ in Eq.(\ref{eq_evolution_operator}), for deriving $\theta_j$. We rely on the procedure utilized in \cite{Szegedy_qw, google_in_a_quantum_network, pagerank_phase} where a decomposition of $W_{P'}$ is provided, facilitating to determine its spectral decomposition. Given the adjacency matrix $A_c$ of a complete graph $\mathcal{G}_c$ with $n$ nodes, we mark one arbitrary node obtaining the matrix $A'_c$ in Eq.s(\ref{system_transition_matrix_complete_graph}, \ref{C_matrix_complete_graph}). We remove one link from the graph $\mathcal{G}_c$ by replacing symmetrically two non zero elements of $A'_c$, $p_{xy}$ and $p_{yx}$, with zero, renaming the matrix $A'$. By normalizing the columns of $A'$ to $1$, we obtain a column stochastic matrix, renaming it $P'$. We reintroduce the matrix $C$ (Eq.(\ref{C_matrix_complete_graph})):
\begin{gather}
	C_{ij}= \sqrt{P_{ij}'P_{ji}'}
\end{gather}
which is symmetric and thus can be diagonalized. Together with the Swap operator defined as:
\begin{gather}
	S= \sum_{i,j=1}^{n}{\ket{i,j}\bra{j,i}},
\end{gather}
matrix $C$ facilitates finding the spectrum of the operator $W_{P'}$. More specifically, the projection operators defined in Eq.(\ref{eq_projection_operators}) satisfy the following relationship \cite{szegedy_equivalence_coin}:
\begin{gather}
	\Pi_B = S\Pi_A S,
\end{gather}	
since they come from a duplication process and the two registers $\ket{x}$ and $\ket{y}$ are completely symmetrical. Consequently, we can write the operator $W_{P'}$ as:
\begin{gather}\notag
	W_{P'}=\mathcal{R}_B \mathcal{R}_A=\left(2\Pi_B-I_{n^2}\right)\left(2\Pi_A-I_{n^2}\right)=\\ \notag
	=\left(2S\Pi_AS-I_{n^2}\right)\left(2\Pi_A-I_{n^2}\right)=\\
	=S\left(2\Pi_A-I_{n^2}\right)S\left(2\Pi_A-I_{n^2}\right)=\left(S\left(2\Pi_A-I_{n^2}\right)\right)^2= U^2,
\end{gather}
with $U=S(2\Pi_B-I_{n^2})$. The spectral decomposition of $U$ is closely related to that of $C$:
\begin{gather}
	\sigma(U)=\{\pm1,e^{\pm i \arccos(\lambda)}\},
\end{gather}
where $\lambda$ are the eigenvalues of $C$, $C\ket{\lambda}=\lambda\ket{\lambda}$. Thus, once found the eigenvalues of $C$, we can determine the eigenvalues of $U$ and subsequently those of $W_{P'}$.
Once determined $W_{P'}$ eigenvalues, we compare them with the analytical value $e^{2i\theta_2}$ (Eq.(\ref{eq_eigenphases_complete_case})).

Among them, we consider the eigenvalue closer to $e^{2i\theta_2}$ to be related to the dynamical eigenphase $\theta_j$ estimated by the QPE, since it requires the highest precision to be distinguished from $\theta_2$ in the QPE procedure. Subsequently, we compute $\theta_j$ and convert the difference between $\theta_j$ and $\theta_2$ to binary, counting the number of bits $l$ before the first $1$ appears. We recall that $l$ is the estimate of the number of bits needed in the QPE first register to distinguish $\theta_j$ from $\theta_2$. This procedure is repeated for $n$ ranging from $4$ to $300$, removing in each case an arbitrary link from the complete graph. It is worth recalling that we marked only one node in each of the graphs, thus we have only two categories of edges, those connecting the marked node with the unmarked ones and those connecting two unmarked nodes.
\begin{figure}
	\centering
	\includegraphics[width=1\columnwidth]{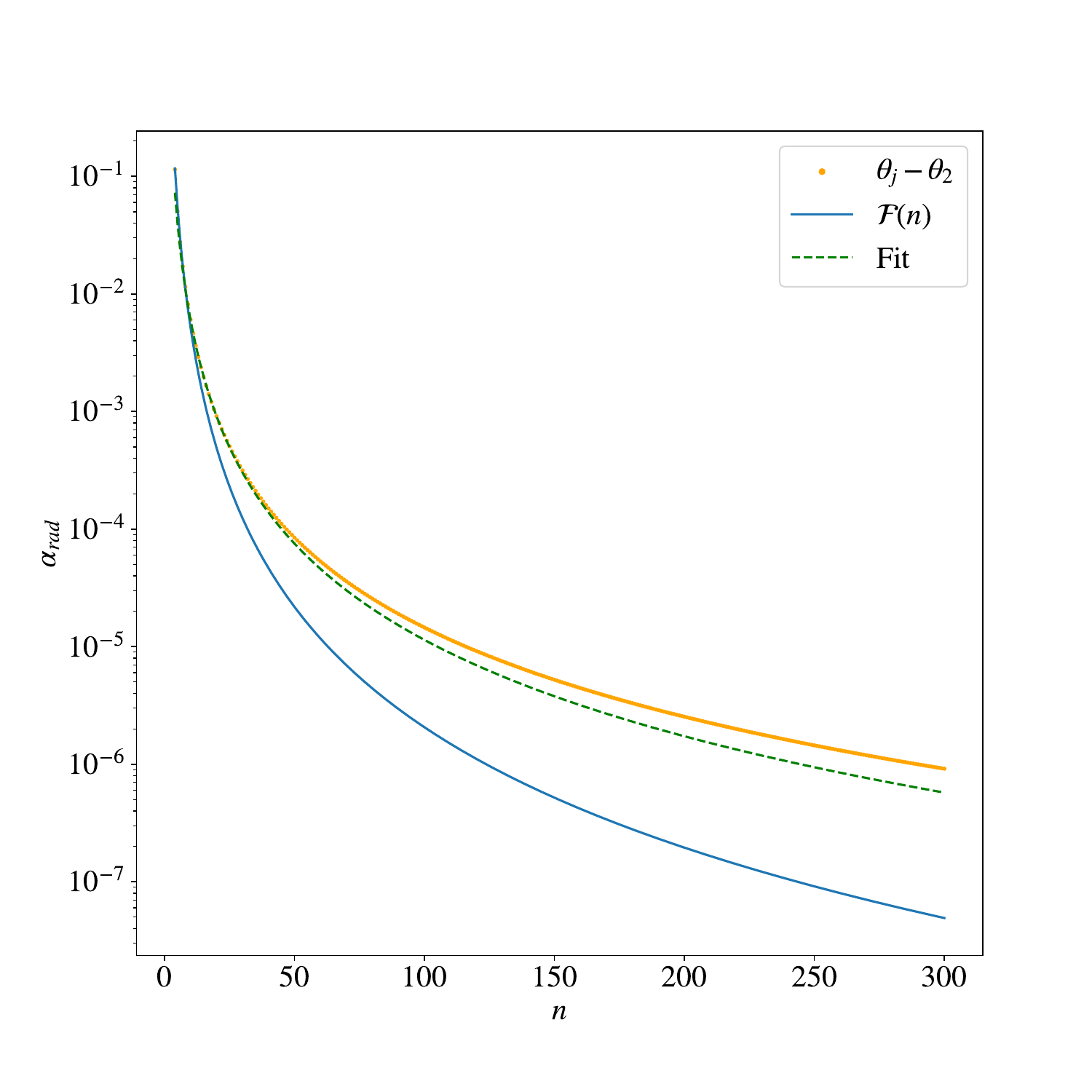}
	\caption{Lower bound $\mathcal{F}(n)$ for the simulated data. In this semi-logarithmic plot the orange points represent the simulated data $\theta_j-\theta_2$, while the blue dashed curve is the fitting obtained through the least square method. The fitting function is adjusted to serve as a lower bound $\mathcal{F}(n)$ for the differences $\theta_j-\theta_2$ with respect to $n$ and it is showed in the plot as a solid blue curve.}
	\label{fit_lower_bound}
\end{figure}

Meanwhile, once determined the values of $\theta_j$ for different $n$, we estimate the number of bits $f$ by fitting the differences $\theta_j-\theta_2$ (Eq.(\ref{eq_thetaj_-_theta2})) with respect to $n$ and subsequently adjusting the fitting function such that it serves as a lower bound for the data, obtaining the function $\mathcal{F}(n)$ in Eq.(\ref{eq_F(n)}). We utilized the least square method for the fitting, obtaining a nonlinear polynomial function. The results are shown in Fig.s \ref{fig_f_vs_l}, \ref{fig_f_vs_l_over_n} and \ref{fit_lower_bound}. In the first two figures we compare the number of bits $l$ obtained from the simulation with the number of bits $f$ obtained from the fitting function $\mathcal{F}(n)$, while in Fig.\ref{fit_lower_bound} we show the lower bound function $\mathcal{F}(n)$, the fitting function and the simulated data.

 \bibliographystyle{unsrt} 
 \bibliography{bibliography}
\end{document}